\documentclass[prx,letterpaper,nobalancelastpage,twocolumn,superscriptaddress,nofootinbib]{revtex4-2}

\usepackage{lineno}
\usepackage{graphicx}
\usepackage{amsmath}
\usepackage{bbold}
\usepackage{amssymb}
\usepackage[english]{babel}
\usepackage{color}
\usepackage[version=4]{mhchem}
\usepackage[hidelinks]{hyperref}
\usepackage{dsfont}
\usepackage{placeins}
\usepackage{mathtools}
\usepackage[normalem]{ulem}
\usepackage{lipsum}
\setcitestyle{super}

\usepackage{soul}
\newcommand{\FCZ}{$0.9962(3)$\,}

\newcommand{\ket}[1]{| #1 \rangle}

\newcommand{\Caltech}{California Institute of Technology, Pasadena, CA 91125, USA}

\newcommand{\Stanford}{Department of Electrical Engineering, Stanford University, Stanford, CA 94305, USA}

\usepackage{caption}
\usepackage[caption=false]{subfig}

\DeclareCaptionLabelSeparator{bar}{ \textbf{\textbar}~}
\usepackage{enumitem}
\setlist{nolistsep}
\begin{document}
\title{Universal quantum operations and ancilla-based readout for tweezer clocks }
 \author{Ran Finkelstein}\thanks{These authors contributed equally to this work}
\author{Richard Bing-Shiun Tsai}
 \thanks{These authors contributed equally to this work}
\author{Xiangkai Sun}
 \thanks{These authors contributed equally to this work}
\author{Pascal Scholl}
\affiliation{\Caltech}
\author{\\Su Direkci}
\affiliation{\Caltech}
\author{Tuvia Gefen}
\affiliation{\Caltech}
\author{Joonhee Choi}
\affiliation{\Caltech}
\affiliation{\Stanford}
\author{Adam L. Shaw}
\affiliation{\Caltech}
\author{Manuel Endres}\email{mendres@caltech.edu}
\affiliation{\Caltech}

\maketitle
\textbf{Enhancing the precision of measurements by harnessing entanglement is a long-sought goal in the field of quantum metrology~\cite{Giovannetti2006,Pezze2018}. Yet attaining the best sensitivity allowed by quantum theory in the presence of noise is an outstanding challenge, requiring optimal probe-state generation and readout strategies~\cite{Huelga1997,Kessler2014A,Rosenband2013,Kaubruegger2021,Marciniak2022}. Neutral atom optical clocks~\cite{Ludlow2015}, leading systems for measuring time, have shown recent progress in terms of entanglement generation~\cite{Eckner2023,Robinson2024,Pedrozo2020}, but currently lack the control capabilities to realize such schemes. Here we show universal quantum operations and ancilla-based readout for ultranarrow optical transitions of neutral atoms. Our demonstration in a tweezer clock platform~\cite{Norcia2019, Madjarov2019,Young2020, Schine2022,Eckner2023, Shaw2024} enables a circuit-based approach to quantum metrology with neutral atom optical clocks. To this end, we demonstrate two-qubit entangling gates with $99.62(3)\%$ fidelity -- averaged over symmetric input states -- via Rydberg interactions~\cite{Evered2023,Ma2023,Schine2022} and dynamical connectivity~\cite{Bluvstein2022} for optical clock qubits, which we combine with local addressing~\cite{Shaw2024} to implement universally programmable quantum circuits. Using this approach, we generate a near-optimal entangled probe state~\cite{Giovannetti2006,Kessler2014A}, a cascade of Greenberger-Horne-Zeilinger (GHZ) states of different sizes, and perform dual-quadrature~\cite{Rosenband2013} GHZ readout. We also show repeated fast phase detection with non-destructive conditional reset of clock qubits and minimal dead time between repetitions by implementing ancilla-based quantum logic spectroscopy~\cite{Schmidt2005} (QLS) for neutral atoms. Finally, we extend this to multi-qubit parity checks and measurement-based, heralded, Bell state preparation~\cite{Lee2022,Verresen2021,Quantinuum2023,Bluvstein2024}. Our work lays the foundation for hybrid processor-clock devices with neutral atoms and more generally points to a future of practical applications for quantum processors linked with quantum sensors\cite{Degen2017}.}

\vspace{1mm}
\begin{figure}[ht!]
	\centering
	\includegraphics[width=\columnwidth]{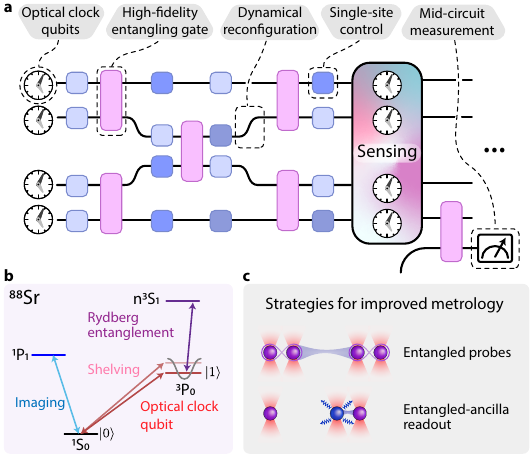}
 \caption{\textbf{Universal quantum operations for tweezer clocks.} \textbf{a}, We demonstrate a universal quantum processor based on optical clock qubits. Such a processor is designed to perform any computational task and at the same time is a highly sensitive clock. To realize such a device, we demonstrate high-fidelity entangling gates between optical clock qubits and circuits with dynamical reconfiguration. We combine these with local single-qubit rotations via sub-wavelength atomic shifts and local mid-circuit readout. \textbf{b}, In our experiment, we use tweezer-trapped $^{88}$Sr atoms, where qubits are based on the ultranarrow optical clock transition $^1\text{S}_0\leftrightarrow ^3\text{P}_0$, and entangling operations are realized through transient excitation to a Rydberg state $n^3\text{S}_1$. We further utilize coherent superpositions of quantized motional states in tweezers to shelve protected clock qubits for mid-circuit detection of ancilla qubits. \textbf{c}, We identify two instances where these universal quantum operations can be used in optimizing quantum metrology: entangled state preparation for near-optimal probes and entanglement-assisted readout with ancilla qubits, realizing quantum logic spectroscopy for neutral atoms.}
	\vspace{-0.5cm}
	\label{Fig1}
\end{figure}

Control over quantum systems is evolving rapidly on many fronts. On the one hand, quantum processors -- which use quantum mechanics to perform calculations -- have matured into intermediate-scale systems with high-fidelity entangling operations and universal control capabilities~\cite{Monz2011,Levine2019,Lu2019,Figgatt2019A,Postler2022,Bluvstein2022,Graham2022,Bravyi2024,Acharya2023}. On the other hand, quantum sensors~\cite{Degen2017} and in particular optical atomic clocks~\cite{Ludlow2015} -- which use quantum mechanics to measure time -- have become both more accurate and precise, revealing new phenomena~\cite{Goban2018} and measuring relativistic effects at a laboratory scale~\cite{Bothwell2022, Zheng2022}. It is thus a natural question, and a long-sought goal, how to utilize quantum information processing tools and entanglement to improve quantum sensors.

Two main strategies are typically considered in this context~\cite{Giovannetti2006}: entangled state preparation and entanglement-assisted measurements. For the first, entanglement introduces correlations between probes, resulting in jointly enhanced sensitivity; unfortunately, this also entails jointly enhanced vulnerability to errors and reduced dynamic range, often nullifying or strongly reducing any effective gains~\cite{Huelga1997,Macieszczak2014,Kessler2014A,Andre2004,Chauvet2010}. Schemes put forward to circumvent this apparent drawback use specific, finely-tuned entangled probe-state preparation and readout schemes~\cite{Giovannetti2006,Kessler2014A,Kaubruegger2021}. While recently one such scheme has been realized with trapped ions~\cite{Marciniak2022}, there is so far no demonstration in neutral atom optical clocks, despite recent progress in entanglement generation~\cite{Eckner2023,Robinson2024,Pedrozo2020}. 

Second, entanglement with an ancillary system can be harnessed in non-destructive readout of a sensor to benefit metrology, \emph{e.g.}, logic spectroscopy of inaccessible probes as demonstrated with ions~\cite{Schmidt2005}, sequential multi-round phase estimation~\cite{Giovannetti2006}, fast mid-circuit detection for multi-ensemble metrology~\cite{Rosenband2013}, and shallow-circuit preparation of long-range entangled states~\cite{Lee2022,Verresen2021,Quantinuum2023,Bluvstein2024}. Only very recently, ancilla-based readout for hyperfine qubits in neutral atoms has been shown~\cite{Bluvstein2024,Anand2024}, but a demonstration for ultranarrow optical transitions -- as needed for optical atomic clocks -- is outstanding. Further, it is critical that ancilla-based readout can be performed repeatedly to fully realize its potential for metrology (and for quantum error correction~\cite{Terhal2015}), which is an outstanding challenge for neutral atoms in general. 

To demonstrate such complex entangled probe-state generation and repeated ancilla-based readout, we realize a scalable universal quantum processor with neutral atom qubits encoded in an ultranarrow optical transition (Fig.~\ref{Fig1}a). Importantly, the same experimental system can be utilized as a neutral atom optical clock~\cite{Ludlow2015}, which currently forms the most stable frequency reference~\cite{Bothwell2022, Zheng2022}. 
Our system specifically is a \textit{tweezer optical clock}~\cite{Norcia2019, Madjarov2019,Young2020, Schine2022,Eckner2023, Shaw2024}, composed of neutral strontium-88 trapped in a tweezer array~\cite{Cooper2018} (Fig.~\ref{Fig1}b,c), which we now equip with universal quantum computing and repeated ancilla-based readout capabilities.

In particular, we demonstrate high-fidelity entangling gates~\cite{Levine2019,Evered2023,Ma2023} for optical clock qubits~\cite{Schine2022}, achieving a state-of-the-art controlled-Z (CZ) fidelity for neutral atoms of \FCZ (averaged over two-qubit symmetric states), and generate an array of optical-clock-transition Bell pairs with a record fidelity.
We further show that optical coherence is preserved during qubit transport in tweezers, previously realized with hyperfine qubits~\cite{Bluvstein2022,Beugnon2007}, enabling dynamical circuit reconfiguration. Together with high-fidelity global single-qubit rotations, single-site control~\cite{Shaw2024} and mid-circuit readout~\cite{Scholl2023B}, this allows us to take a circuit-based approach to realize strategies for entangled state preparation and readout designed to improve metrology (Fig.~\ref{Fig1}c). 

\vspace{3mm}
\noindent\textbf{High-fidelity entangling gates for optical qubits}
\noindent We begin by demonstrating high-fidelity entangling gates for the ultranarrow optical clock transition. The optical transition defines a qubit composed of the ground state of strontium-88 $\mathrm{^1S_0}$, denoted as $|0\rangle$, and the metastable state $\mathrm{^3P_0}$, denoted as $|1\rangle$ (Fig.~\ref{Fig1}b). To realize high-fidelity entangling operation, we apply a single phase-modulated Rydberg pulse for a CZ gate, which was shown to be time-optimal~\cite{Pagano2022,Jandura2022} and recently implemented in hyperfine and nuclear atomic qubits~\cite{Evered2023,Ma2023}. To simultaneously realize high-fidelity single-qubit rotations, we operate at a magnetic field of 450 G which balances between sufficiently high Rabi frequency on the clock transition (scaling linearly with the magnetic field) and sufficiently strong interactions (as the Rydberg interaction strength is magnetic field dependent, see Methods, Ext. Data Fig.~\ref{SI_exp_requirement}). We prepare atoms in the $\ket{1}$ state and perform erasure-cooling~\cite{Scholl2023B} which prepares atoms close to the motional ground state.
\begin{figure}[t!]
	\centering
	\includegraphics[width=\columnwidth]{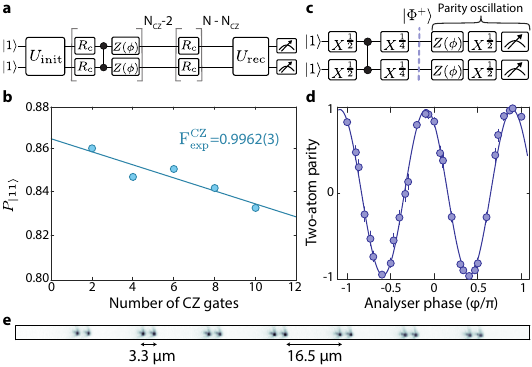}
	\caption{\textbf{High-fidelity entangling gates for optical clock qubits.} \textbf{a}, Randomized circuit characterization of CZ gate fidelity averaged over symmetric input states prepared by $\hat{U}_{\mathrm{init}}$. The circuit includes a fixed number of random single-qubit gates $R_C$ and a variable number $\mathrm{N_{CZ}}$ of CZ gates. A final unitary $\hat{U}_{\mathrm{rec}}$ is precomputed to maximize the return probability $P_{|11\rangle}$ in the absence of errors. \textbf{b}, By fitting the return probability as a function of $\mathrm{N_{CZ}}$, we infer a gate fidelity of \FCZ after correcting for false contribution from leakage errors (Methods).
 \textbf{c}, A circuit for creating and characterizing the Bell state $\ket{\Phi^+}$. \textbf{d}, Parity oscillation with a measured bare parity contrast of $0.963^{+7}_{-10}$. Together with the population overlap (Methods), we obtain a Bell state generation fidelity of $0.976^{+4}_{-6}$ ($0.989^{+4}_{-6}$ SPAM corrected). In \textbf{b},\textbf{d}, data is conditioned on atom pairs surviving erasure cooling, which occurs at a rate of 51$\%$.
 \textbf{e}, Array configuration depicted by average atomic fluorescence image. The interatomic separation is chosen to induce strong Rydberg interactions within each pair and avoid unwanted interactions between pairs.} 
	\vspace{-0.5cm}
	\label{Fig_gates}
\end{figure}

\begin{figure*}[ht!]
	\centering
	\includegraphics[width=\textwidth]{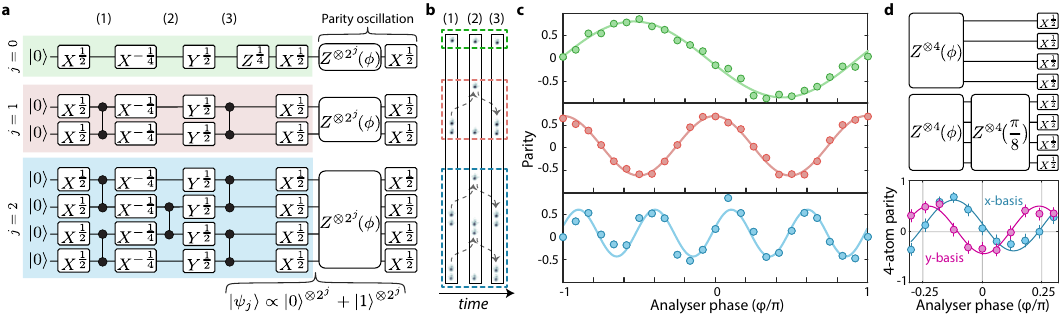}
	\caption{\textbf{Simultaneous generation of a cascade of GHZ states and GHZ dual-quadrature readout.} \textbf{a}, A quantum circuit for the simultaneous preparation and characterization of GHZ states of the form $\ket{\psi_j}\propto(\ket{0}^{\otimes 2^j}+\ket{1}^{\otimes 2^j})$, here $j=0,1,2$. \textbf{b}, Array reconfiguration steps depicted by average atomic fluorescence images at each one-dimensional array configuration (1), (2), (3), as marked in the circuit in \textbf{a}. Dashed arrows represent atom movements performed between two sequential array configurations. \textbf{c}, Simultaneous parity oscillations for $j=0,1,2$, from top to bottom, showing a 1:$1.96(6)$:$3.9(1)$ ratio in phase sensitivity. \textbf{d}, Simultaneous dual-quadrature parity readout of two copies of a 4-atom GHZ state realized through a $\pi/8$ phase shift on each atom in one copy, resulting in a collective $\pi/2$ phase shift before the final readout pulse.}
	\vspace{-0.5cm}
	\label{Fig_GHZ}
\end{figure*}
We characterize the performance of such gates via a randomized circuit (Fig.~\ref{Fig_gates}a) designed to benchmark the CZ gate fidelity averaged over the two-qubit symmetric states (Methods, Ext. Data Fig.~\ref{SI_RRB}).
Varying the number of CZ gates $\mathrm{N_{CZ}}$, we infer a CZ fidelity of \FCZ by fitting the decay of the return probability as a function of $\mathrm{N_{CZ}}$ to an exponential trend (Fig.~\ref{Fig_gates}b). The quoted fidelity is corrected for false contribution from leakage out of the qubit manifold during the gate (Methods). We numerically find this gate benchmark to be within $5\%$ of the infidelity averaged over Haar random states for our gate parameters~\cite{FutureWork}.
Our result sets a new state-of-the-art fidelity in Rydberg-based two-qubit entangling gates for long-lived qubits~\cite{Evered2023, Peper2024}.

As an important application of the CZ gate, we generate an array of Bell pairs with optical clock qubits (Fig.~\ref{Fig_gates}c). After initializing atoms in superposition states and applying a CZ-gate, a $\pi/4$ pulse rotates the state into the even-parity, metrologically relevant, Bell state $\ket{\Phi^+}=(1/\sqrt{2})(|00\rangle+|11\rangle)$. For the state obtained in the experiment, we measure the population overlap with the ideal Bell state as well as drive parity oscillations (Fig.~\ref{Fig_gates}d) to infer the off-diagonal (coherence) terms~\cite{Sackett2000A}. Overall, we obtain a Bell state generation fidelity of $0.976^{+4}_{-6}$ ($0.989^{+4}_{-6}$ SPAM corrected) which is limited, beyond the CZ fidelity, by the finite fidelity of clock $\pi/2$ pulses and decoherence during the circuit (Methods). 

\vspace{3mm}

\noindent\textbf{Simultaneous GHZ cascade generation and dual-quadrature readout}\newline
\noindent Having demonstrated high-fidelity gates on optical clock qubits, we now turn to preparing larger entangled states and executing algorithms requiring more complex circuits.
Specifically, we first demonstrate the preparation of a complex quantum register with metrological significance based on a cascade of entangled GHZ states. Such GHZ states enable, in principle, the best sensitivity for linear observables allowed by quantum theory - the Heisenberg limit - with precision scaling as $1/N$, where $N$ is the atom number. However, for a single $N$-qubit GHZ state, while the sensitivity to phase variations is increased by a factor of $N$, the dynamic range of phases that can be unambiguously measured is likewise \textit{reduced} by a factor of $N$~\cite{Huelga1997,Kessler2014A,Macieszczak2014}. A similar trade-off exists for spin-squeezed entangled states~\cite{Andre2004,Chauvet2010}.
Importantly, a smaller dynamic range means a shorter allowable interrogation time, resulting in precision loss.

A proposed scheme~\cite{Giovannetti2006,Kessler2014A} to circumvent this trade-off between sensitivity and dynamic range requires the simultaneous preparation of a cascade of $M$ different groups of GHZ-like states of the form $\ket{\psi_j}\propto(\ket{0}^{\otimes 2^j}+\ket{1}^{\otimes 2^j})$ with $j=0,1...M-1$. Such a probe state enables Heisenberg-limited sensitivity, up to logarithmic corrections, while preserving the dynamic range. It thus enables a viable path to quantum-enhanced metrology under realistic conditions where local oscillator noise is the main limitation, as in current neutral atom optical clocks~\cite{Macieszczak2014}. We note that these proposals are quantum-enhanced extensions of the classical optimal interrogation schemes~\cite{Rosenband2013,Borregaard2013} that were recently experimentally shown~\cite{Shaw2024,Zheng2024}.

We consider different circuit realizations for generating the required simultaneous preparation of GHZ states with $j=0,1,2$, utilizing our universal quantum processor aided by dynamical array reconfiguration. While large GHZ states can be prepared with circuits applying extensive local control (Ext. Data Fig.~\ref{SI_circuit_GHZ8}), here we realize a circuit compiled with global control up to its final layer (Fig.~\ref{Fig_GHZ}a). This choice is guided by minimizing idle time added by local operations, during which laser-atom dephasing occurs. The local control for single-qubit rotations is achieved through local phase shifts imparted via sub-wavelength shifts in atomic position~\cite{Shaw2024}, and the connectivity for two-qubit gates is realized by mid-circuit array reconfiguration (Fig.~\ref{Fig_GHZ}b). We verify the simultaneous preparation of increasingly phase-sensitive states by varying the phase of a final $\pi/2$ pulse and measuring the parity signal $\mathcal{\hat{P}}=\prod_i{\hat{Z}_i}$ for each GHZ state. In Fig.~\ref{Fig_GHZ}c, we plot this signal against phase and find a 1:$1.96(6)$:$3.9(1)$ ratio in phase sensitivity, consistent with the 1:2:4 ratio expected.

This demonstration measures only one quadrature of the optical phase signal. Importantly, the dynamic range can be doubled if both phase quadratures are probed simultaneously, so-called dual-quadrature readout~\cite{Kessler2014A, Rosenband2013,Shaw2024,Zheng2024}.
We demonstrate such dual-quadrature readout for entangled states (Fig.~\ref{Fig_GHZ}d) by preparing two 4-atom GHZ states and then imparting a phase rotation to only one through local phase shifts. This enables readout in the ${\hat{Y}}$-basis \textit{in parallel} to readout in the ${\hat{X}}$-basis. Here, a $\pi/8$ phase shift per atom ($\pi/2$ phase on the entangled 4-atom register) is realized by moving~\cite{Shaw2024} all four atoms by $\lambda/16\approx 44 $ nm. We note that such phase shifts could alternatively be realized by moving a single atom in the GHZ state by $\lambda/4$.  

An important question concerns the limitations on the maximum attainable GHZ state size in terms of achievable fidelity. We hypothesize that the phase noise of our clock laser is the dominant error source, which we test with experimental and numerical investigations. In the simultaneous GHZ state preparation scheme, we achieve a bare parity contrast of $0.52(3)$ for $N$=4 (Fig.~\ref{Fig_GHZ}c). To reduce the impact of clock laser phase noise, we prepare $N$=4 and $N$=2 GHZ states using a circuit with shorter idle time, \emph{i.e.} wait time during which operations are done on other registers (by removing the single site operation $\hat{Z}^{\frac{1}{4}}$ shown in Fig.~\ref{Fig_GHZ}a). With this sequence, we find an improved contrast of $0.68(3)$ and fidelity of $0.71(2)$ (Ext. Data Fig.~\ref{SI_ZNE_data}). We note that the contrasts from both sequences are at or above the threshold of $1/\sqrt{N}=0.5$ required for a metrological gain~\cite{Colombo2022,Pezze2018} at a fixed, short interrogation time. However, they do not exceed the higher threshold for the cascaded-GHZ scheme (Methods).

Motivated by this, we study fidelity as a function of the circuit idle time. We measure a decay in fidelity and a modification of the bitstring distribution (Ext. Data~Fig.~\ref{SI_ZNE_data}) which follows the prediction of our error model, consistent with our hypothesis that clock laser phase noise is the dominant error source. We further run a model using a frequency noise spectrum of a 26-$\text{mHz}$ clock laser system~\cite{Bishop2013}, for which we predict a $N$=4 GHZ state fidelity $\approx$ 0.97 (Methods, Ext. Data Fig.~\ref{SI_sim_better_clock}). Thus, we expect significantly improved results with an upgraded clock laser system, surpassing the metrological gain threshold for the cascaded-GHZ scheme, even for larger $N$ (Methods, Ext. Data Fig.~\ref{SI_metrological_gain}).
\begin{figure}[t!]
	\centering
	\includegraphics[width=\columnwidth]{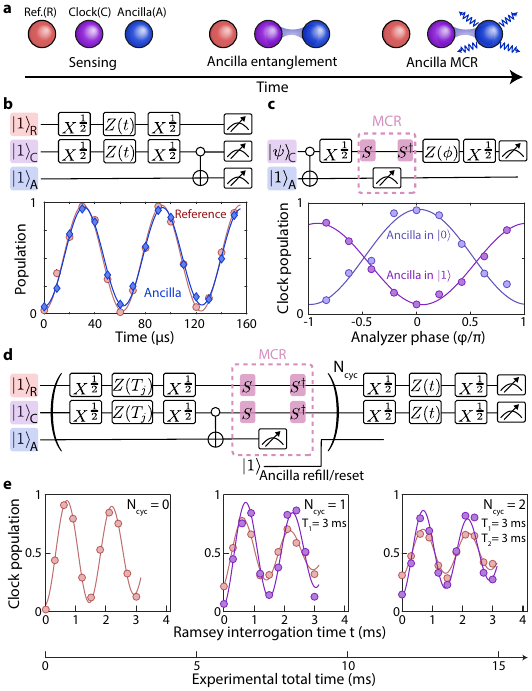}
	\caption{\textbf{Repeated ancilla-based quantum logic spectroscopy (QLS).} \textbf{a}, We consider three ensembles of atoms: reference atoms (R; red), clock atoms (C; purple), and ancillas (A; blue). An entangling gate maps the state of clock atoms to the state of ancilla atoms before mid-circuit readout (MCR) of the ancilla state. \textbf{b}, Logic spectroscopy via ancilla atoms. A Ramsey interrogation is performed in parallel on reference atoms (R) and clock atoms (C). A clock-ancilla CNOT gate (controlled on qubit state $|0\rangle$) maps the state onto ancilla atoms (A) for readout. \textbf{c}, Ancilla-based $\hat{X}$-measurement. A non-destructive measurement of the state $|\psi\rangle_C$ is realized by shelving clock qubits ($\hat{S}$ operation) followed by ancilla MCR. We show that following unshelving ($\hat{S}^\dag$), the state of the clock qubit is projected to a well-defined state in the measurement basis, conditioned on the ancilla state. \textbf{d}, $\rm{N_{cyc}}$ repeated rounds of QLS with minimal (2.9 $\text{ms}$) dead time, enabled by ancilla replacement via array reconfiguration. Each round $j$ consists of a fixed evolution time $T_j$ followed by ancilla-based $\hat{X}$ measurement before a final round where the clock qubit is directly measured in a Ramsey sequence with variable time t. \textbf{e}, We find improved contrast in the final measurement round when conditioned on ancilla results (see text) in all previous rounds (purple markers) as compared to reference clock atoms which are protected during MCR but do not interact with ancillas (red markers).
	} 
	\vspace{-0.5cm}
	\label{Fig_ancilla_single}
\end{figure}

\vspace{3mm}
\noindent\textbf{Ancilla-based quantum logic spectroscopy}\newline
Having used entanglement to \textit{prepare} metrological probe states, we now turn to use it to \textit{read out} the resulting state. Specifically, we demonstrate a method based on mapping quantum information from a set of sensor qubits to a set of ancillary qubits, which are measured while the sensor qubits are quickly available for sensing once more.
Such \textit{repeated}, non-destructive, ancilla-based measurements, are a central building block in quantum error detection and correction~\cite{Terhal2015}.
In a metrological context, repeated mid-evolution measurement and reset cycles are desirable for efficient multi-ensemble schemes~\cite{Rosenband2013}. These schemes enable extended dynamic range and effective coherence time, even in the absence of entangled probes. However, fast mid-circuit measurement and reset of atomic qubits are challenging due to the overhead in resetting both the motional state and the electronic state of a qubit after direct measurement.

To efficiently realize ancilla-based readout in atomic systems, information-carrying data qubits must first be protected from global imaging light and their coherence must be preserved during mid-circuit readout (MCR) of ancilla qubits. Such protection can be realized by using dual-species arrays~\cite{Anand2024}, cavity-assisted readout~\cite{Deist2022}, a separate readout zone~\cite{Bluvstein2024}, or shelving to states dark to imaging light~\cite{Graham2023,Lis2023A,Norcia2023A}. In our system, we take the last approach and have recently realized MCR~\cite{Scholl2023B} by shelving optical qubits into superpositions of motional states which are long-lived and dark to imaging light.

By performing high-fidelity entangling gates between information-carrying (clock) qubits and ancilla qubits before MCR (Fig.~\ref{Fig_ancilla_single}a), we can realize a form of quantum logic spectroscopy (QLS) with neutral atom optical clock qubits, as originally introduced for mixed-species ion traps~\cite{Schmidt2005}.
In our case, we use local control to prepare certain atoms in superposition states sensitive to the clock laser detuning $\Delta$, while ancilla atoms remain in the insensitive state $|1\rangle$. Clock qubits acquire a time-dependent phase $\Delta\cdot t$ which is mapped onto population in a Ramsey sequence. A subsequent clock-ancilla CNOT gate (controlled on qubit state $|0\rangle$) then maps this signal onto the ancilla state (Fig.~\ref{Fig_ancilla_single}b). 
This sequence realizes an ancilla-based detection of $\hat{X}$ of the clock qubit.

To realize a non-destructive ancilla-based $\hat{X}$-measurement, we then shelve clock qubits and read out the ancilla state. To this end, we ramp up the depth of ancilla traps by a factor of 5 (leaving traps with clock atoms unchanged) such that they are non-resonant with the global motional shelving drive. Following shelving of clock qubits, we perform ancilla MCR of the signal field~$\Delta$. 
Conditioned on the ancilla measurement result, the clock atom is projected to a known state in the measurement basis, which enables repeated interrogation. We verify this by measuring the clock qubit state in the $\hat{X}$-basis via a $\pi/2$ pulse with a variable phase $\phi$ before detection (Fig.~\ref{Fig_ancilla_single}c). We find the clock qubit is projected onto the $|+\rangle$ ($|-\rangle$) state conditioned on the ancilla measurement result $|0\rangle$ ($|1\rangle$) in the $\hat{Z}$-basis, with a contrast of $0.84(4)$ ($0.73(4)$). These results are conditioned on successful one-way shelving~\cite{Scholl2023B,Ma2023} (Methods).   

\begin{figure*}[t!]
	\centering
	\includegraphics[width=\textwidth]{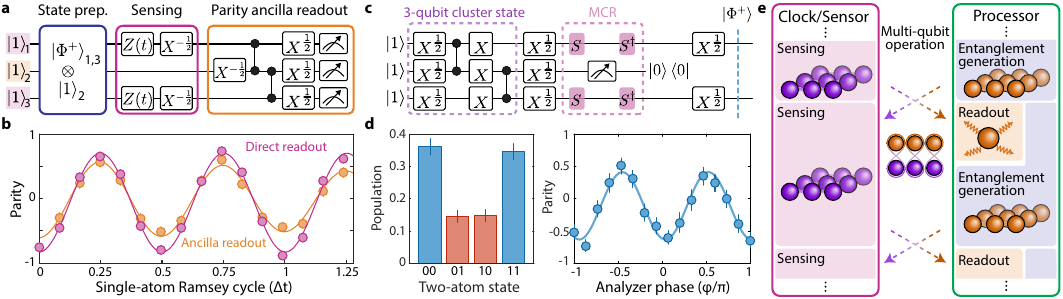}
	\caption{\textbf{Ancilla parity readout and measurement-based Bell state generation.} \textbf{a}, An ancilla qubit (middle) is prepared in parallel with a Bell pair, which is then used in Ramsey interrogation. A parity signal is measured either directly on the Bell pair or through ancilla-based parity readout. \textbf{b}, Direct readout of the Bell pair parity signal (pink) is compared with a single ancilla readout (orange) yielding a weight-2 parity signal oscillating at the same frequency and phase within error bars (Methods). \textbf{c}, A circuit for measurement-based preparation of a long-range Bell pair, composed of 3-qubit cluster state generation followed by MCR on the middle ancilla qubit. Conditioning on the ancilla result $|0\rangle$ ($\approx 45\%$ of experimental shots) projects the two exterior atoms to the Bell state $\ket{\Phi^+}$. \textbf{d}, Characterization of the resulting Bell state with population overlap of $0.71(2)$ in the even parity subspace $\{\ket{00}, \ket{11}\}$. The parity oscillation of the resulting state yields a contrast of $0.52(3)$, resulting in an overall raw Bell state fidelity of $0.61(2)$. Data is conditioned on the successful shelving of the two exterior qubits ($\approx 82\%$ of experimental shots). \textbf{e}, A proposal for a continuous clock based on a modular architecture (see Outlook for details). 
	} 
	\vspace{-0.5cm}
	\label{Fig_ancilla_parity}
\end{figure*} 
Ideally, non-destructive readout enables repetitive measurements of the same qubit or the same register. This can be achieved by resetting the ancilla qubit and repeating the readout cycle. In our case, we utilize an 18 $\mu$s fast imaging method~\cite{Scholl2023A} which is much shorter than the qubit coherence time but results in a high probability of losing the ancilla atom during imaging. We thus realize repeated rounds of measurement and reset by replacing the ancilla atom with a reserve atom, which was shelved during ancilla MCR, using dynamical reconfiguration. This yields a dead time of only 2.9 $\text{ms}$ dominated by shelving and unshelving operations. Alternatively, we demonstrate reusing the same ancilla atom by applying a 10-ms low-loss imaging followed by re-cooling (Methods, Ext. Data Fig.~\ref{SI_ancilla_reset}).

We utilize ancilla-based readout to perform fast and repeated phase detection ($\hat{X}$-measurement) cycles with clock qubit reset (Fig.~\ref{Fig_ancilla_single}d). In Fig.~\ref{Fig_ancilla_single}e, we plot the resulting Ramsey signal for up to three consecutive measurement rounds. We use information from ancilla detection in previous rounds to remove shots where the ancilla state was toggled from $|1\rangle$ to $|0\rangle$ (maintaining $78\%$ of the experimental shots for a single round of ancilla detection), indicating a phase flip. We contrast this with reference qubits that did not interact with ancillas and find an improved contrast for the ancilla-based detection. For the latter, the contrast is now limited by the fidelity of shelving and unshelving operations as part of the MCR. 

While here ancillas are used to measure a single qubit, extending our demonstration to a larger ensemble of ancillas interacting with an ensemble of clock qubits would enable single-shot phase estimation of the global field $\Delta$. Information on phase deviation outside of the dynamic range could be used to correct the phase estimation of the longer evolving ensemble realized by the reference qubits in our demonstration~\cite{Rosenband2013}. Such fast detection and reset are further advantageous for low dead-time clock operation, reducing the instability due to the Dick effect~\cite{Ludlow2015}. On the other hand, using ancillas in a single-species array comes at the expense of a reduced number of clock qubits, increasing quantum projection noise, such that the metrological gain depends on the balance of the two. 

\vspace{3mm}

\noindent\textbf{Multi-qubit parity checks}\newline
We now generalize our experimental demonstration of ancilla-based readout to multi-qubit measurements. In a multi-qubit measurement, an ancilla is entangled with multiple qubits and then measured.
The measurement projects the register qubits to a well-defined parity subspace. This may prove especially useful for the preparation of large-scale long-range entangled states, such as GHZ states, through ancilla measurements~\cite{Lee2022,Verresen2021,Quantinuum2023,Bluvstein2024}, for which we show a proof-of-principle for optical qubits. We note that multi-qubit measurements, measuring higher weight observables, are also central to stabilizer-based quantum error correction~\cite{Terhal2015}.  

We demonstrate the lowest-order version of such schemes, a weight-2 parity measurement, where the ancilla is entangled with two different clock qubits before measurement. To do this, we prepare an input state composed of an ancilla atom in the clock state $|1\rangle$ and two atoms in a Bell state and perform a $\hat{X}\hat{X}$ parity measurement (Fig.~\ref{Fig_ancilla_parity}a).
In Fig.~\ref{Fig_ancilla_parity}b, we compare the direct readout of a parity signal from the Bell pair (pink) with the ancilla readout (orange). We find that indeed the single ancilla atom observable now oscillates at a frequency of twice the single atom detuning $\Delta$. This frequency, as well as the phase of the signal, is consistent with a direct readout of the Bell pair parity (Methods). 

We now turn to demonstrate another relevant application of the conditional parity projection in a weight-2 ancilla MCR: long-range entanglement induced by measurement~\cite{Lee2022,Verresen2021,Quantinuum2023,Bluvstein2024}. We first prepare a 3-qubit cluster state and then perform MCR of the central qubit in the $\hat{X}$-basis (Fig.~\ref{Fig_ancilla_parity}c). As $\hat{Z}\hat{X}\hat{Z}$ is a stabilizer of the cluster state, this measurement generates a Bell pair with a well-defined parity for the exterior atoms. Conditioning on the ancilla result $|0\rangle$, we find the Bell state $\ket{\Phi^{+}}$ with population $0.71(2)$ in the even parity subspace $\{\ket{00}, \ket{11}\}$ and a parity oscillation contrast of $0.52(3)$, yielding an overall raw fidelity of $0.61(2)$ (Fig.~\ref{Fig_ancilla_parity}d). If instead we condition on the ancilla result $|1\rangle$, we find the state is projected to the odd parity Bell state $\ket{\Psi^{-}}=(1/\sqrt{2})(\ket{01}-\ket{10})$, which is not metrologically useful in our context, with $0.64(2)$ in the odd parity subspace $\{\ket{01}, \ket{10}\}$. Alternatively, by performing an ancilla-based $\hat{X}\hat{X}$ measurement, one can prepare both metrologically relevant Bell states $\ket{\Phi^{+}}$,$\mbox{$\ket{\Phi^{-}}=(1/\sqrt{2})(\ket{00}-\ket{11})$}$. Such schemes can be considered a heralded preparation, which could be made deterministic by a feedback operation~\cite{Lee2022,Verresen2021,Quantinuum2023,Bluvstein2024}.
We note that with improvements in fidelity, and with sufficiently fast feedback, this could then be readily extended to perform measurement-based preparation of large-scale GHZ states with only a constant depth circuit~\cite{Verresen2021,Lee2022,Quantinuum2023}. This is in contrast to linear or log-depth standard unitary circuits~\cite{Graham2022}, which could be instrumental for generating large-scale optical clock GHZ states in future devices. 

\vspace{3mm}
\noindent\textbf{Outlook}\newline~
Our results demonstrate how universal quantum processing and ancilla-based readout can be integrated with a neutral atom optical clock. Extending such schemes to larger two-dimensional tweezer clock systems~\cite{Young2020} would enable the creation of multiple copies of GHZ states at a given size. We note in particular that the schemes demonstrated here could be easily generalized to two dimensions by performing all operations in parallel for a set of rows to generate multiple copies of GHZ states.
We see this as a viable avenue towards quantum-enhanced operation of neutral atom clocks using cascades of GHZ states~\cite{Giovannetti2006,Kessler2014A} combined with classically optimal interrogation schemes and quadrature readout~\cite{Rosenband2013,Borregaard2013, Shaw2024,Zheng2024}. 

We have demonstrated a CZ fidelity of \FCZ, averaged over the two-qubit symmetric states, setting a new state-of-the-art for entangling gate fidelity, as well as the full toolbox needed to execute relevant algorithms for metrology and computation with optical clock qubits. The main limitation to current circuit fidelities is the clock laser phase noise resulting in dephasing during circuit idle times. The ancilla-based MCR is further limited by the fidelity of shelving qubits in motional states.  
Upgrading the clock laser system used in this work to match the current state-of-the-art~\cite{Bothwell2022,Young2020} would enable generating significantly larger GHZ states than demonstrated here (Ext. Data Fig.~\ref{SI_sim_better_clock}, Methods). We then expect the maximum GHZ state size will eventually be limited by a combination of clock laser phase noise and CZ gate fidelity. In this context, we note that the CZ gate fidelity is currently dominated by phase noise of the Rydberg laser (Ext. Data Fig.~\ref{SI_extended_2QRB}, Methods), which could easily be improved by implementing cavity filtering techniques. Further, utilizing native, multi-qubit gates~\cite{Cao2024,Evered2023} or effective all-to-all connectivity~\cite{Bluvstein2022} could reduce the required circuit depth. Ancilla-based MCR would also improve with a higher stability clock laser and better control and homogenization of trap waists~\cite{Scholl2023B}. Finally, GHZ states can be prepared with constant depth circuits via measurement-based schemes~\cite{Lee2022,Verresen2021}, for which we showed first steps with optical clock qubits in Fig.~\ref{Fig_ancilla_parity}c,d. 

Our results further point to a modular device with distinct quantum processing and sensing modules, realizing a vision of a quantum computer connected to a high-precision sensor (Fig.~\ref{Fig_ancilla_parity}e). The processing module could be used to prepare entangled probe states \textit{in parallel} to sensing, followed by a state exchange operation between the sensor and the processor. Subsequently, readout could be performed in the processing module in parallel to sensing, before preparing new entangled probe states, yielding very low dead time operation. 
Options for fully realizing this vision could be based on a multi-zone approach with a single species~\cite{Bluvstein2024}, a dual-species~\cite{Anand2024} device composed of alkaline-earth and alkali atoms (\emph{e.g.}, strontium and rubidium), an optical-metastable-ground-state qubit architecture~\cite{Lis2023A,Chen2022} or combinations thereof. 
Several of our results can be viewed as proof-of-principle demonstrations towards this modular vision, including state exchange operations (Ext. Data Fig.~\ref{SISwap_Bell}) and higher-weight ancilla-based readout. Such a device could be part of next-generation quantum technologies from field-transportable sensors to state-of-the-art optical clocks~\cite{Bothwell2022} in the quest to push the limits of precision measurements.

\vspace{3mm}
\noindent\textbf{Acknowledgements}\newline
We acknowledge fruitful discussions with Chi Zhang and Kon Leung. We acknowledge support from the Army Research Office MURI program (W911NF2010136), the NSF QLCI program (2016245), from the Institute for Quantum Information and Matter, an NSF Physics Frontiers Center (NSF Grant PHY-1733907), the NSF CAREER award (1753386), and the DARPA ONISQ program (W911NF2010021). Support is also acknowledged from the U.S. Department of Energy, Office of Science, National Quantum Information Science Research Centers, Quantum Systems Accelerator. RF acknowledges support from the Troesh postdoctoral fellowship. RBST acknowledges support from the Taiwan-Caltech Fellowship. XS acknowledges support from the Todd Alworth Larson Fellowship. TG acknowledges support from the Quantum Science and Technology Scholarship of the Israel Council for Higher Education. JC acknowledges the support from the Terman Faculty Fellowship at Stanford.

\vspace{3mm}
\noindent\textbf{Data availability}\newline
The data supporting this study's findings are available from the corresponding author upon reasonable request.

\vspace{3mm}
\noindent\textbf{Code availability}\newline
The codes supporting this study's findings are available from the corresponding author upon reasonable request.

\vspace{3mm}
\noindent\textbf{Competing interests}\newline
The authors declare no competing interests.

\vspace{3mm}
\noindent\textbf{Author contributions}\newline
R.F., R.B.T., and X.S. performed the experiments, data analysis, and numerical simulations. S.D., R.F., and T.G. performed the metrological gain calculation and analysis. R.F., R.B.T., X.S., P.S., J.C, and A.L.S. contributed to the experimental setup. M.E. supervised this project. All authors contributed to writing the manuscript.

\FloatBarrier

\newpage
\bibliography{library_endreslab}
\bibliographystyle{adamref}


\newpage
\clearpage
\setcounter{equation}{0}  
\setcounter{figure}{0}  
\captionsetup[figure]{labelfont={bf},name={Ext. Data Fig.},labelsep=bar,justification=raggedright,font=small}
\section*{Methods}

\subsection*{Experimental considerations and constraints}
Our experimental setup has been detailed in previous work~\cite{Madjarov2019, Scholl2023A, Shaw2024}. We discuss experimental requirements specific to executing quantum circuits on a quantum processor with optical clock qubits. In particular, several trade-offs need to be balanced in the choice of magnetic field, trap depth, and inter-atomic spacing.

\begin{figure*}[]
    \centering
    \includegraphics[width=\textwidth]{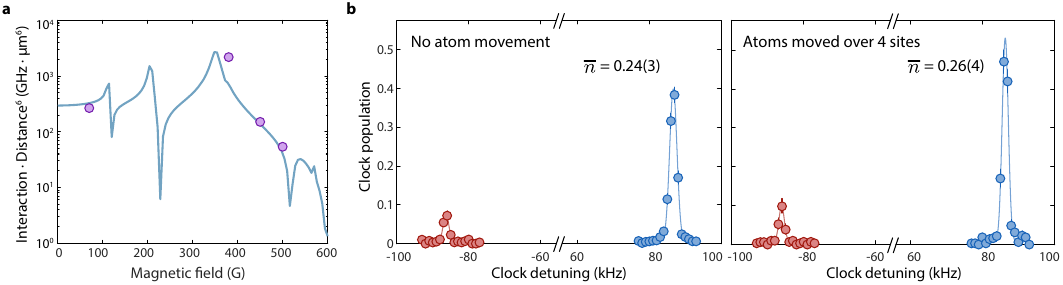}
    \caption{\textbf{Rydberg interaction strength and minimal heating due to dynamical reconfiguration.} \textbf{a}, Rydberg interaction strength at different magnetic fields. The solid line represents numerical calculations performed for a 7 $\mathrm{\mu m}$ spacing with the 'PairInteraction' package~\cite{Weber2017}. Markers represent interaction energy measured via a two-atom spin echo experiment at $\approx$ 7 $\mathrm{\mu m}$ spacing. \textbf{b}, No observable heating due to dynamical array reconfiguration. An array reconfiguration step consists of transporting the atoms over multiple inter-tweezer distances to desired positions. We use sideband spectroscopy to measure the average radial motional occupation $\Bar{n}$ in the tweezers before and after such atom movements. We move the atoms over 13 $\mu$m in 160 $\mu$s, the typical distance and duration in all of the experiments presented in the main text. We perform this measurement at nominal temperature $\Bar{n} = 0.24(3)$ with no further erasure-cooling, as opposed to the temperature $\Bar{n} = 0.01$ with erasure-cooling~\cite{Scholl2023B}. After performing dynamical reconfiguration, the measured $\Bar{n}$ is 0.26(4), \emph{i.e.} no significant heating (within error bars), an important feature for high-fidelity rotations of optical clock qubits. } 
    \label{SI_exp_requirement}
\end{figure*}

Due to laser frequency noise, high-fidelity single-qubit rotations benefit from large Rabi frequency on the clock transition ($^1\text{S}_0$ $\leftrightarrow$ $^3\text{P}_0$). The clock transition Rabi frequency scales linearly with the magnetic field and we achieve $\Omega=2\pi\cdot2.1$ kHz at 450 G. On the other hand, the Rydberg interaction strength varies with the magnetic field due to admixing with other Rydberg states~\cite{Weber2017}. Specifically, a numerical calculation (using the 'Pairinteraction' package~\cite{Weber2017} and limiting the considered Rydberg states to $n\pm5$ for faster convergence) shows that the interaction energy peaks around 380 G and decreases for higher magnetic fields (Ext. Data Fig.~\ref{SI_exp_requirement}a). Our experimental measurements at several magnetic fields are consistent with this overall trend. Therefore, we operate at a magnetic field of 450 G which balances between sufficiently high clock Rabi frequency and sufficiently strong Rydberg interactions. 

Our nominal tweezer trap depth ($U_0 \approx 450 \mu$K) has been chosen to perform efficient atom loading and high-survival, high-fidelity imaging~\cite{Covey2019A}, as well as efficient driving of the carrier transition for clock qubits. However, we find that the Rydberg gate performs slightly better when the trap is turned off. In our case, this is predominantly due to beating between adjacent tweezers, which results in trap depth fluctuations at a frequency of 650 kHz (equal to the tone separation on our tweezer-creating AOD). As the Rydberg transition ($^3\text{P}_0$ $\leftrightarrow$ $61^3\text{S}_1$) is not under magic trap conditions, this results in detuning noise for the gate operation. For an optimized CZ gate with traps kept on at $0.2U_0$, we find a two-qubit gate fidelity lower by $\approx3\cdot10^{-3}$ (not shown). By adding an AOM on the tweezer optical path, we implement a fast switch-off of trapping light (rise/fall time on the order of 50 ns). At this timescale, we find that switching traps off from a more shallow trap depth $0.2U_0$ is preferable as it imparts minimal heating and no observed loss for clock qubits. Furthermore, in mid-circuit readout (MCR), efficient motional shelving relies on strong sideband coupling~\cite{Scholl2023B}, which is stronger as the Lamb-Dicke factor increases (trap depth decreases). On the other hand, array reconfiguration benefits from sufficiently deep traps. 

The experimental sequence thus involves adiabatic ramps of trap depth as well as fast switch-off and -on. After loading the atoms into the tweezers at $U_0$, we lower the tweezer depth to $0.5U_0$ to perform erasure-cooling~\cite{Scholl2023B}. For gate operations, we drive coherent clock rotations at $0.2U_0$ and switch the trap off for about 500~ns to perform the Rydberg entangling pulse. When selective local mid-circuit readout is applied, we adiabatically ramp to deeper traps of $U_0$ for the ancilla qubits, while holding the clock qubits at fixed depth.

We now discuss how we perform dynamical array reconfiguration, shown in Fig.~\ref{Fig1} as part of a full quantum operation toolbox. We coherently transport an atom across multiple sites by performing a minimal-jerk trajectory that follows $x(t)=6 t^5-15 t^4+10 t^3$ for $t\in [0,1]$. For this trajectory, the acceleration is zero at the two endpoints, which avoids the sudden jump in the acceleration profile and minimizes the associated jerk, aiming for minimal heating, which is especially important for driving optical clock transitions in the sideband-resolved regime.

With this trajectory, we find no significant temperature increase for atoms that we transport over 4 sites (equivalent to 13.26 $\mu$m) in 160 $\mu$s at trap depth $U_0$ (Ext. Data Fig.~\ref{SI_exp_requirement}b). This is the typical distance applied for dynamical array reconfiguration. Another aspect of this inter-atomic spacing choice is 13.26 $\mu$m $\approx$ 19$\times$698 nm (corresponding to the clock transition wavelength), ensuring an effective zero displacement-induced phase shift~\cite{Shaw2024}.

\subsection*{Single-qubit (clock) error model}
\begin{figure*}[]
	\centering
	\includegraphics[width=\textwidth]{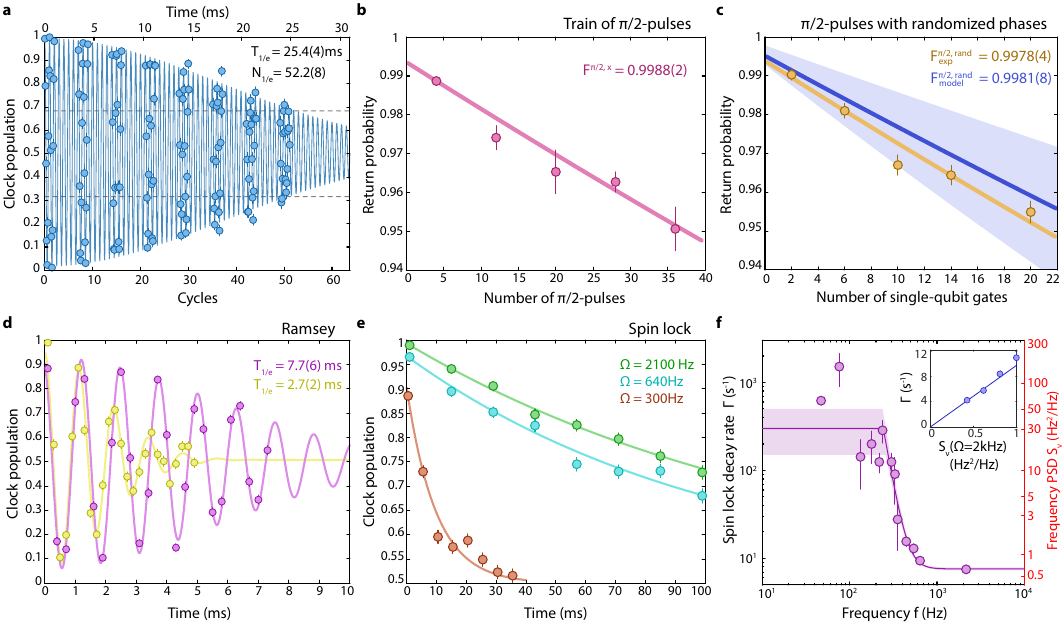}
	\caption{ \textbf{Verification of clock error model and clock laser frequency noise power spectral density (PSD).} \textbf{a}, Rabi oscillations between $\ket{0}$ and $\ket{1}$. When averaging over the entire array, the Rabi oscillation shows a coherence time of $25.4(4)~\mathrm{ms}$ and $52.2(8)$ coherent cycles (dashed lines represent $1/e$ oscillation amplitude). The long-term coherence is mainly limited by clock laser frequency noise and Rabi frequency inhomogeneity across the array. \textbf{b}, The return probability of driving a train of $\pi/2$ pulses with the same phase. We obtain a per-pulse fidelity of $0.9988(2)$ from this method. \textbf{c}, Random phase $\pi/2$ pulse benchmarking. We randomly draw $N$ $\pi/2$ pulses from the set $\{R_X (\pi/2), R_{-X} (\pi/2), R_Y (\pi/2), R_{-Y} (\pi/2)\}$ under the constraint that the final state returns to the initial state with perfect single-qubit operations. We obtain a single-qubit $\pi/2$ fidelity of $0.9978(4)$ (yellow) from this method. The described single-qubit (clock) error model (Methods) predicts the $\pi/2$ fidelity being $0.9981(8)$ (blue), the error bar of which indicates the model parameter uncertainty (shaded area). Note that all single-qubit operations are performed globally. \textbf{d}, Ramsey measurements taken on two different days, representing the typical fluctuations of the clock laser performance. We use these to estimate the low-frequency part of the clock laser frequency PSD. \textbf{e}, The spin-lock experiment at different Rabi frequencies. We fit the clock population as a function of time using an exponential function with time constant $\tau$ and determine the decay rate $\Gamma = 1/\tau$ at each frequency. \textbf{f}, Measured decay rates at different frequencies and the inferred laser frequency power spectral density. We find that the spin-lock decay rate is linearly proportional to the PSD at the corresponding Rabi frequency (Inset). We interpolate the PSD down to low frequencies (solid line). The shaded area represents the model uncertainty, considering the fitting model uncertainty and the observed Ramsey coherence time fluctuations as in \textbf{d}. 
	} 
	\label{SIClock_error}
\end{figure*}

We characterize our ability to perform coherent single-qubit rotations with a global addressing beam and test our error model by driving the clock transition on atoms with an average motional occupation of $\Bar{n}\approx 0.01$, following erasure-cooling along the optical axis~\cite{Scholl2023B}. We drive Rabi oscillations with a nominal Rabi frequency $\Omega = 2.1~\mathrm{kHz}$ and observe $52.2(8)$ coherent cycles (Ext. Data Fig.~\ref{SIClock_error}a). Applying a train of $\pi/2$ pulses along the $X$-axis, we find a per-pulse fidelity of $0.9988(2)$. We note that in such sequences the effect of slow frequency variations is suppressed. We thus characterize the $\pi/2$ pulse fidelity by applying a pulse train of $\pi/2$ pulses with random rotation axes $\pm X,\pm Y$~\cite{Knill1998}. The resulting $\pi/2$ pulse fidelity is measured to be $0.9978(4)$ (Ext. Data Fig.~\ref{SIClock_error}c).

The dominant error source in the single-qubit operation is laser frequency noise, which is characterized by the frequency power spectral density (PSD) function $S_\nu(f)$. We characterize this by Ramsey and spin-lock~\cite{Bodey2019} sequences.

\textit{Ramsey.--} The Ramsey sequence is sensitive to the low-frequency component of the laser frequency PSD up to the inverse of the Ramsey interrogation time ($\approx 100~\mathrm{Hz}$). In our experimental setup, we observe day-to-day fluctuations in the Ramsey coherence time (Ext. Data Fig.~\ref{SIClock_error}d). We use an effective model of the PSD at low frequencies (Ext. Data Fig.~\ref{SIClock_error}f) to account for the fluctuations of the Ramsey coherence time. We set the PSD to be a constant $H$ at low frequencies up to some frequency of interest ($\approx 200~\mathrm{Hz}$) and numerically find that the Ramsey coherence time is inversely proportional to $H$.

\textit{Spin lock.--}
To probe and quantify fast frequency noise up to our Rabi frequency, we perform a spin-lock sequence. We initialize all atoms in an eigenstate of $\hat{X}$ and turn on a continuous drive along the X axis for a variable time. Then we apply a $\pi/2$ pulse along the Y axis, which transfers all atoms into state $\ket{1}$ in the absence of errors. The probability of returning to $\ket{1}$ decays over time (Ext. Data Fig.~\ref{SIClock_error}e), and the decay rate is predominantly sensitive to frequency noise at this Rabi frequency~\cite{Bodey2019}. By varying the Rabi frequency of the continuous drive field and measuring the decay of the probability in $\ket{1}$, we determine the frequency PSD utilizing the linear relation between the decay rate and the frequency PSD $S_\nu(f)$ at the Rabi frequency (Ext. Data Fig.~\ref{SIClock_error}f).

To account for both the fast frequency noise measured by the spin-lock experiment and the slow frequency noise that determines the Ramsey coherence time, we interpolate the laser frequency PSD with a power-law function $S_\nu(f) = h_0 + (h_\alpha / f)^\alpha$ upper bounded by $H$ at low frequencies. The model parameters $h_0, h_\alpha,$ and $\alpha$ are obtained from fitting the spin-lock data (Ext. Data Fig.~\ref{SIClock_error}e, f). The upper bound $H$ is flexible within a range (shown as the shaded area in Ext. Data Fig.~\ref{SIClock_error}f) to effectively describe the day-to-day Ramsey coherence time fluctuations. This range is reflected in the uncertainties of the error model predictions quoted throughout this work.

In addition to laser frequency noise, we note that although the single-qubit operations are sensitive to finite temperature, we perform erasure-cooling~\cite{Scholl2023B} to prepare atoms close to their motional ground state ($\Bar{n}\approx0.01$). This leads to a negligible impact ($\approx 1\cdot 10^{-4}$) on the clock $\pi/2$ pulse fidelity as predicted by our error model. 

In addition to the error sources described above, we also include laser intensity noise, pulse shape imperfection, spatial Rabi frequency inhomogeneity, and Raman scattering induced by the tweezer light. All these error sources result in an aggregate of $\approx 1\cdot 10^{-4}$ infidelity for the clock $\pi/2$ pulse.

With all described error sources included, the error model predicts an average $\pi/2$ fidelity of $0.9981(8)$ (Ext. Data Fig.~\ref{SIClock_error}c), in good agreement with the experimental value of 0.9978(4).

\subsection*{Two-qubit gate fidelity benchmarking}
\begin{figure*}
    \centering
    \includegraphics{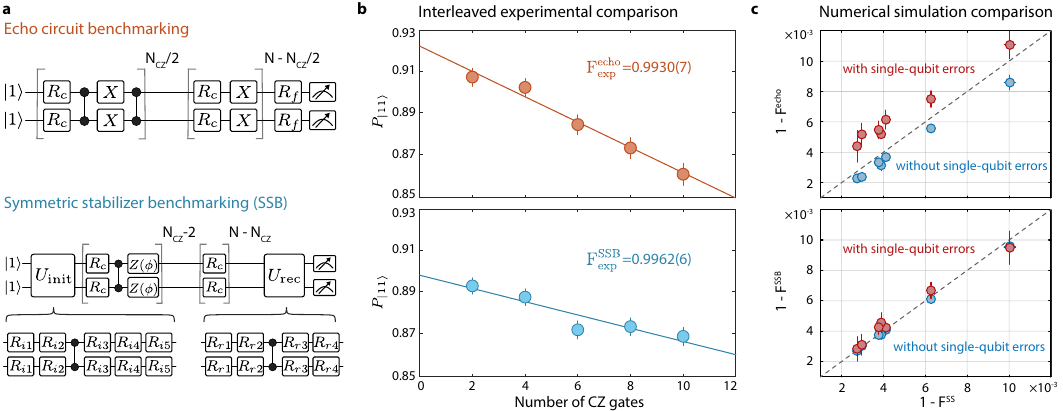}
    \caption{\textbf{Comparison of benchmarking circuits.} \textbf{a,} Top: a randomized circuit where CZ gates are interleaved with global single-qubit $\pi$-pulses (echo pulses) and random $\pi/2$ rotations $R_c$ (echo circuit). Bottom: a randomized circuit in which an initial unitary $\hat{U}_{init}$ prepares one of the 12 two-qubit symmetric stabilizer states with a uniform probability via a series of single-qubit rotations $R_i$ and a CZ gate. The circuit (SSB circuit) is then designed to maintain such uniform distribution, in order to reduce sensitivity to single-qubit gate errors (see Methods). Each Rydberg CZ gate is composed of a Rydberg pulse followed by a phase gate $Z(\phi)$ which we apply virtually by shifting the phase of the next global rotation. $Z(\phi)$ is separately calibrated to find the single-atom phase acquired during the Rydberg pulse. 
    A final unitary $\hat{U}_{rec}$ returns the atoms to the state $|11\rangle$ in the absence of errors. This unitary is also composed of single-qubit rotations $R_r$ and a CZ gate.
    \textbf{b,} Interleaved experimental comparison of the two randomized circuits described in a. Using the same gate we find a difference of about $3\times10^{-3}$ between benchmarking methods, stemming from the additional sensitivity of the echo circuit to single-qubit gate errors. This is consistent with the results of a full error model accounting for clock and Rydberg imperfections. \textbf{c,} We use a numerical simulation of the full error model to compare the true fidelity overlap averaged over the symmetric stabilizer states $F^{\text{SS}}$ with the inferred fidelity from the echo circuit benchmarking $F^{\text{echo}}$ (top) and with the inferred fidelity from the symmetric stabilizer circuit benchmarking $F^{\text{SSB}}$ (bottom). By rescaling the magnitude of different error sources, we find the SSB circuit to be a good proxy of the CZ fidelity averaged over symmetric stabilizer states. We further find the sensitivity of this sequence to single-qubit gate errors to be significantly attenuated relative to the echo benchmarking circuit (Methods).  
    }
    \label{SI_RRB}
\end{figure*}

Here, we give more details on the randomized circuit (shown in Fig.~\ref{Fig_gates}), which is used to benchmark the CZ gate fidelity.
We have first applied a randomized circuit similar to the one proposed and used in Ref.~\cite{Evered2023}, which includes echo pulses ($\pi$-pulses along $X$) interleaved with random single-qubit rotations and CZ gates (Ext. Data Fig.~\ref{SI_RRB}a). For this circuit, we observe that both two-qubit (Rydberg) errors and single-qubit (clock) errors contribute to the inferred infidelity (Ext. Data Fig.~\ref{SI_RRB}b,c). The fact that such a circuit is sensitive to single-qubit gate errors, although the number of single-qubit gates is kept fixed, stems from the fact that the probability distribution of two-qubit states before each single-qubit gate changes as a function of the number of CZ gates applied. Note that as errors affect entangled states and non-entangled states differently, changing the probability distribution would result in a non-unity return probability even if the fidelity of CZ gates is perfect, in the presence of single-qubit gate errors.
In this context, the sequence used in Extended Data Fig. 6 of Ref.~\cite{Evered2023} also shows sensitivity to single-qubit errors, as the probability distribution between entangled and non-entangled states is not fixed as a function of number of CZ gates.   

To mitigate this effect, we have designed a randomized circuit (Ext. Data Fig.~\ref{SI_RRB}a) where the probability of finding any one of the 12 two-qubit symmetric stabilizer states is maintained uniform, irrespective of the number of CZ gates, at each stage of the circuit~\cite{FutureWork}. We term this circuit symmetric stabilizer benchmarking (SSB).
Specifically, the probabilities of finding an entangled or separable state are equal throughout the circuit.   

Using an interleaved experimental comparison we find a difference of about $3 \cdot 10^{-3}$ between benchmarking methods in the fidelity directly inferred from the slope of the return probability (Ext. Data Fig.~\ref{SI_RRB}b). This difference stems from the additional sensitivity of the echo circuit benchmarking to single-qubit gate errors. 
This observation is in good agreement with a full error model, accounting for both clock and Rydberg excitation imperfections (Ext. Data Fig.~\ref{SI_RRB}c). This model further confirms that fidelity inferred from the SSB circuit is an accurate proxy of the gate fidelity averaged over all two-qubit symmetric stabilizer states. 
We confirm this observation holds over a wide range of error rates by rescaling the strength of individual error sources in the numerical model (Ext. Data Fig.~\ref{SI_RRB}c). These include incoherent and coherent errors. However, we note that coherent errors or gate miscalibration of larger magnitude would result in an increased error in estimating gate fidelity, which is a common issue across various benchmarking techniques. We further note that the gate fidelity averaged over all two-qubit symmetric stabilizer states is equal to the gate fidelity averaged over two-qubit symmetric input states. This can be seen from the fact that these symmetric stabilizer states form a quantum state two-design on the symmetric subspace~\cite{FutureWork}.

\subsubsection*{Correcting for false contribution from leakage errors}
We readout the return probability for the randomized circuit benchmarking by pushing out ground-state atoms and pumping clock-state atoms to the ground-state for imaging. As part of this optical pumping, any population in the $^3\mathrm{P}_2$ state would be pumped and identified as \emph{bright}, \emph{i.e. }clock state population. 
We thus further correct for the effect of leakage from the Rydberg state into the state $^3P_2$ identified as \emph{bright}. We separately measure the decay into $^3P_2$ per gate by repeating the benchmarking sequence followed by push out of atoms in the qubit subspace and repumping the $^3P_2$ state for imaging. At Rydberg Rabi frequency of 5.4 MHz, the false contribution to the CZ fidelity is measured to be $1.8(4) \cdot 10^{-4}$ per gate, in good agreement with numerical predictions. CZ fidelity quoted throughout this work has been corrected downward for this effect.

\subsubsection*{Two-qubit gate (Rydberg) error model}
\begin{figure*}[]
    \centering
    \includegraphics[width=\textwidth]{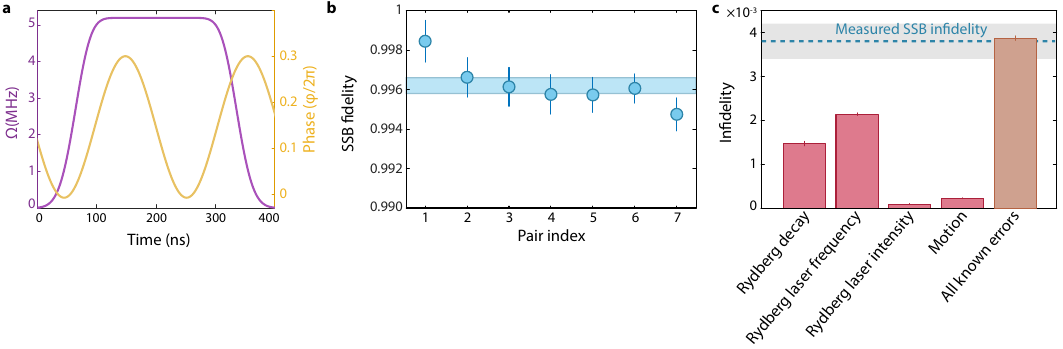}
    \caption{\textbf{Controlled-Z gate via Rydberg interactions and associated error budget.} \textbf{a}, Rydberg pulse shape that realizes a CZ gate between two optical qubits. We adopt the time-optimal pulse~\cite{Jandura2022,Evered2023,Ma2023} with a Rydberg Rabi frequency of 5.4 MHz. We optimize the parameters of the phase modulation function and the Rydberg detuning by varying each one in search of the maximal return probability as part of the randomized circuit used to benchmark the fidelity (Fig.~\ref{Fig_gates}a). 
    \textbf{b}, Pair-resolved fidelity, inferred from the sequence in Fig.~\ref{Fig_gates}a. The blue shaded area represents the array-averaged fidelity 0.9962(4). \textbf{c}, Our error model predicts the infidelity to be 0.0387(5), consistent with the experimental result of 0.0038(4) (blue dashed line and shaded area). 
    } 
    \label{SI_extended_2QRB}
\end{figure*}

\begin{figure*}[]
	\centering
	\includegraphics[]{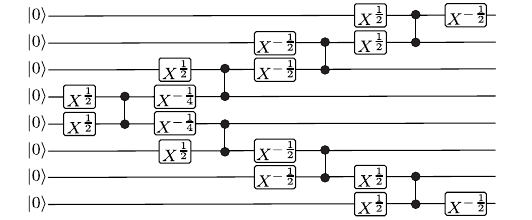}
	\caption{\textbf{Quantum circuit for generating 8-qubit GHZ state.} With the toolbox proposed in this work, one can in principle prepare arbitrary-size GHZ states, limited by the number of atoms. We give a circuit that generates an 8-qubit GHZ state while respecting the 1D array constraint in our current setup.  
	} 

	\label{SI_circuit_GHZ8}
\end{figure*}

\begin{figure*}[]
	\centering
	\includegraphics[width=\textwidth]{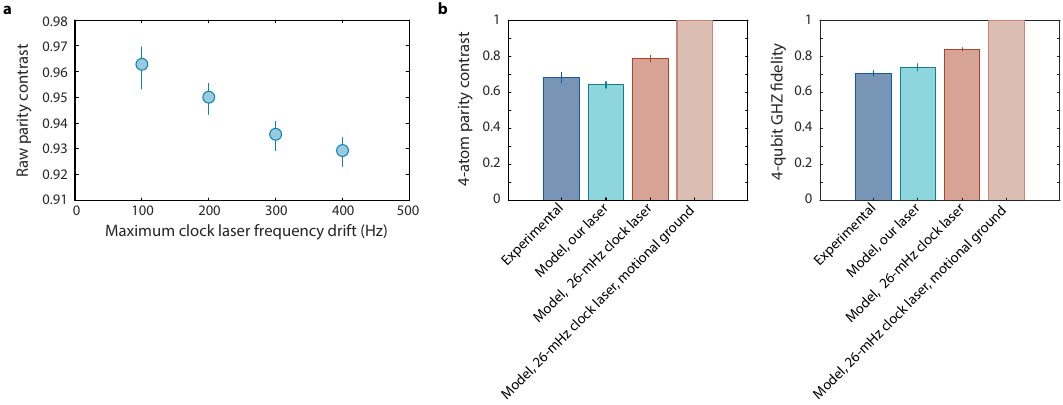}
	\caption{\textbf{Effects of clock frequency noise on entangled state fidelity.} \textbf{a}, Experimental Bell state raw parity contrast as a function of maximum clock laser frequency drift, determined by clock laser feedback frequency (Methods). We analyze the Bell state parity experimental runs with associated clock feedback frequencies less than a certain maximum drift cutoff. The fitted raw parity contrast shows a clear decreasing trend as we increase the cutoff frequency. The parity contrast shown in Fig.~\ref{Fig_gates}d is taken with the cutoff frequency of 100 Hz. \textbf{b}, Experimental results and error model prediction for the 4-qubit GHZ state parity contrast (left) and fidelity (right). In our experimental sequence, the 4-qubit GHZ state generation fidelity is mainly limited by clock laser frequency noise and finite atomic temperature. We simulate the circuit with our error model including the clock laser frequency PSD and finite temperature of $\Bar{n}=0.24$. We also assume perfect CZ gates. The error model predicts the parity oscillation contrast to be 0.66 and the state fidelity to be 0.75, which are consistent with our experimental realization (contrast being 0.68(3) and fidelity being 0.71(2)). We also simulate the same circuit with our clock error model with the frequency PSD of a 26-mHz laser\cite{Bishop2013}. The model predicts a contrast of 0.79 and a fidelity of 0.84 for $\Bar{n}=0.24$. With zero temperature ($\Bar{n}=0$), the clock error model predicts near-unity state fidelity ($>0.999$). The limitation in this regime is the entangling gate fidelity. With the high-fidelity entangling gate presented in this work, we estimate a 4-qubit GHZ state generation fidelity of $\approx0.97$. 
    }
	\label{SI_sim_better_clock}
\end{figure*}

Our model for the two-qubit gate accounting for Rydberg errors is based on previous modeling of errors during Rydberg entangling operations~\cite{Choi2023,Scholl2023A}. We adapt it to model the dynamics of a three-level system with ground ($\ket{0}$), clock ($\ket{1}$), and Rydberg state ($\ket{r}$). Following the optimization of the gate parameters for a time-optimal pulse~\cite{Evered2023,Ma2023} in the error-free case (Ext. Data Fig.~\ref{SI_extended_2QRB}a), we fix these parameters and simulate noisy dynamics with the Monte Carlo wavefunction approach. The model includes Rydberg laser intensity noise, Rydberg laser frequency noise, Rydberg decay (quantum jumps), and atomic motion. The predicted contribution of each error source to the CZ gate infidelity is shown in Ext. Data Fig.~\ref{SI_extended_2QRB}c.
For the analysis shown in Fig. \ref{SI_RRB}c, we repeat the numerical simulation several times and change the magnitude of one of the error model parameters in each run. For example, we rescale the overall magnitude of the noise PSD for frequency or intensity noise or the Rydberg decay rate.

\subsection*{Data-taking and analysis}
\subsubsection*{Data-taking and clock laser feedback}
We discuss the general data-taking procedure for all experiments shown in the main text. Typically, each experimental repetition takes $\approx$ 1 second. To collect enough statistics, we perform the same sequence for several hours up to several days (for the randomized circuit two-qubit gate characterization). However, on this timescale, the clock laser reference cavity experiences environmental fluctuations, resulting in clock laser frequency drifts from $\approx$ 10 Hz to $\approx$ 100 Hz on the time scale of $\approx$ 10 minutes.

We thus interleave data-taking with calibration and feedback runs~\cite{Choi2023, Shaw2024}. To measure the clock laser detuning from atomic resonance, we perform Rabi spectroscopy with the same nominal power and $\pi$ pulse time as used in the experiment. The laser frequency is then shifted accordingly by an AOM. Such feedback is performed every 5-10 minutes, depending on the details of the experimental sequence. We further record the applied laser frequency shifts, which can serve as an indicator of the clock laser stability during the experimental runs. To compare the stability from experiment to experiment, we take the standard deviation of the feedback values. In the main text, the gate benchmarking (Fig.~\ref{Fig_gates}a,b) and the simultaneous preparation of a cascade of GHZ states (Fig.~\ref{Fig_GHZ}) had a feedback standard deviation of 73 Hz and 68 Hz, respectively. During the data-taking of the optical-clock-transition Bell state generation experiment (Fig.~\ref{Fig_gates}c,d), the feedback standard deviation was 203 Hz, significantly higher than other experiments. To ensure consistency of clock laser conditions among all experiments, we select the Bell state generation experimental runs with associated clock laser feedbacks less than 100 Hz. After applying this cutoff, the standard deviation of the feedback frequencies is 67 Hz, comparable with the other experiments.

To study the effects of the short-term clock stability, we analyze the Bell state parity experimental runs with associated clock feedback frequencies less than a certain cutoff. With the cutoff frequency increasing from 100 Hz (the results presented in Fig.~\ref{Fig_gates}d) to 400 Hz (all data included), the parity contrast shows a clear decreasing trend (Ext. Data Fig.~\ref{SI_sim_better_clock}a). This is consistent with our Bell pair generation fidelity being limited by clock laser phase noise.

In contrast, we note that with the SSB randomized circuit method for characterizing the CZ gate itself, we find results that are consistent run-to-run and day-to-day, within our experimental error bars. This further attests to the largely reduced sensitivity of this sequence to single-qubit gate errors stemming from clock laser drift and clock laser phase noise.

\subsubsection*{Error bars and fitting}
Error bars on individual data points throughout this work represent 68\% confidence intervals for the standard error of the mean. If not visible, error bars are smaller than the markers.


The randomized circuit return probability shown in Fig.~\ref{Fig_gates}b and the parity signal shown in Fig.~\ref{Fig_gates}d are fitted using the maximum-likelihood method~\cite{Scholl2023A} (see details in the next section), and error bars on fitted parameters represent one standard deviation fit errors.
Fitting for all other experimental data is done using the weighted least squares method.

\subsubsection*{Data analysis of Bell state fidelity}
For Bell state experimental results, we analyze the data under the same principle as in our previous work~\cite{Scholl2023A}, where we consider the Beta distribution to assess the underlying probabilities. For the parity signal shown in Fig.~\ref{Fig_gates}d, we fit the data with a sine function with four free parameters: offset, contrast, phase, and frequency. We find that using the maximum-likelihood method taking the underlying Beta distribution of each data point into account is necessary, as the standard Gaussian fit typically \textit{overestimates} the contrast by $\approx0.015$. This is because the Beta distribution deviates from Gaussian distribution when the two-atom parity is close to $\pm 1$, which breaks the underlying assumption of Gaussian fit. From this, we obtain a parity contrast of $0.963^{+7}_{-10}$ ($0.983^{+7}_{-10}$ SPAM corrected). Together with the measured population overlap $P_{00}$+$P_{11}$ = $0.988^{+5}_{-7}$ ($0.994^{+5}_{-7}$ SPAM corrected) (not shown), we obtain Bell state generation fidelity of $0.976^{+4}_{-6}$ ($0.989^{+4}_{-6}$ SPAM corrected). These results are obtained from analyzing experimental runs with associated clock feedback frequencies less than 100 Hz.

\subsubsection*{State preparation and measurement correction}
The dominant measurement error stems from the long tails in a typical fluorescence imaging scheme. In our experiment, we inferred the imaging true negative and true positive being $F_0=0.99997$ and $F_1=0.99995$, respectively, from experimental measurements via a model-free calculation (Ref.~\cite{Madjarov2021} Section 2.6.7). We note that these are not state detection fidelities in a circuit, which requires additional push-out before imaging~\cite{Covey2019A}. The probability of successfully expelling the ground-state atom from the trap for state discrimination is $B=0.9989(1)$. 
Taking these into account, the single-atom measurement-corrected values $P_0^m$ and $P_1^m$ have the following relation with the raw values $P_0^r$ and $P_1^r$:

\begin{equation}
    \begin{bmatrix}
        P_0^m \\ P_1^m
    \end{bmatrix}
    =
    \begin{bmatrix}
        1-C & 1-A-C \\
        C & A+C
    \end{bmatrix}
    \begin{bmatrix}
        P_0^r \\ P_1^r
    \end{bmatrix},
\end{equation}
where $A = [B(F_0+F_1-1)]^{-1} = 1.0012$, $C = 1-F_1[B(F_0+F_1-1)]^{-1} = -0.0011$. Assuming that the measurement is independent among the atoms, we extend this correction to multi-qubit measurements by taking the Kronecker product of the above matrix. 

After measurement correction, we can now correct for state preparation errors for the Bell pair generation circuit (Ext. Data Fig.~\ref{SI_Bell_SPAM}). At the circuit initialization (state preparation) stage, we implement an erasure-cooling scheme~\cite{Scholl2023B} and analyze the results conditioned on no erasure detected. We identify that the dominant imperfections in this state preparation stage are \textit{i)} atom loss (with probability $\varepsilon _l = 0.0027$ for single atom) and \textit{ii)} decay from $\ket{1}$ to $\ket{0}$ (with probability $\varepsilon_d = 0.0037$ for single atom). We keep track of how all two-qubit initial states ($\ket{11}, \ket{10}, \ket{01}, \ket{1,\text{lost}}, \ket{\text{lost}, 1},$ ...) contribute to the population distribution and the coherence at the measurement stage. 

For population distribution, apart from the ideal initial state $\ket{11}$, we keep track of how the erroneous initial states evolve under a perfect circuit execution and contribute to the final population distribution (Ext. Data Fig.~\ref{SI_Bell_SPAM}). We correct the bitstring populations to the first order of $\varepsilon_d$ and $\varepsilon_l$. Following the probability tree, we write:
\begin{equation}
\begin{split}
    P_{00}^m &= (1-2\varepsilon_l-2\varepsilon_d) P_{00}^{c} + \dfrac{1}{4}\cdot 2\varepsilon_d + \cos^2 \dfrac{\pi}{8}\cdot2\varepsilon_l,\\ P_{11}^m  &= (1-2\varepsilon_l-2\varepsilon_d) P_{11}^{c} + \dfrac{1}{4}\cdot2\varepsilon_d,
\end{split}
\end{equation}
where bitstring probabilities $P_b^m$ are measurement-corrected and $P_b^c$ (the SPAM corrected population) are issued from perfect initial state preparation, inherent to the quantum circuit execution errors.

For the coherence measurement, we keep track of how different erroneous initial states contribute to the observed parity contrast. The error channel with one lost atom (initial state being $\ket{1,\text{lost}}, \ket{\text{lost},1}$) does not affect the contrast due to a different oscillation frequency. On the other hand, if an atom has decayed to the ground state ($\ket{01}, \ket{10}$), its parity oscillation frequency remains the same but with a $\pi$-phase shift and a contrast of 0.5. This contributes negatively to the observed parity contrast. Hence, the measured parity oscillation contrast $C^m$ (after measurement correction), in terms of the SPAM corrected contrast $C^c$, to the first order of error probabilities, is
\begin{equation}
    C^m=(1-2\varepsilon_l-2\varepsilon_d)C^c - 2\varepsilon_d\cdot\frac{1}{2}
\end{equation}

\subsubsection*{Erasure conversion for motional qubit initialization}
In the main text, there are several results where the analysis of a final image is conditioned on a preceding fast image which verifies state preparation (after erasure cooling) or motional qubit initialization (after shelving).  
First, we note that erasure cooling is strictly only needed for the motional qubit initialization in mid-circuit measurements. Additionally, we find improved single-qubit (clock) gate fidelities following erasure cooling, and this improvement becomes significant for shallow tweezers.

In contrast, the improvement in CZ gate fidelity following erasure cooling is insignificant. In a full error model we find only a $2\cdot 10^{-5}$ increase when cooling the radial degree of freedom to its motional ground state. 

For experiments where mid-circuit readout is applied (Fig. 4, Fig. 5), we report the results conditioned on not detecting atoms in the ground state after a shelving pulse~\cite{Scholl2023B}. We provide the results here with no conditioning for completeness. For measurement-based Bell state generation (Fig.~\ref{Fig_ancilla_parity}d), without erasure excision, the contrast is 0.39(3), and the population overlap would be 0.64(2), yielding a raw Bell state fidelity of 0.52(2). For ancilla-based $\hat{X}$ measurement (Fig.~\ref{Fig_ancilla_single}c), the contrast, conditioned on ancilla result $\ket{0}$ ($\ket{1}$), is 0.60(3) (0.45(3)).

We attribute the limited shelving fidelity mostly to the limited Rabi frequency on the sideband transition, compared with typical frequency variations of the addressing laser, or the trap frequency. Further limitations may arise from the uniformity of the trap waists (and depths) or different tweezers across the array. These limitations can be overcome with a more stable clock laser, or by employing more advanced pulse sequences designed to be insensitive to such inhomogeneities~\cite{Gong2023A}.

\subsection*{Clock error effects on 4-qubit GHZ state preparation}
As discussed in the main text, during the preparation of the 4-qubit GHZ state, the entangled state is vulnerable to finite atomic temperature and clock laser noise. Entangled states dephase during the array reconfiguration time, which is considered idle time in the quantum circuit, due to laser frequency noise. To quantitatively study the effect of laser frequency noise, we perform the experiment with different idle times. We then measure the parity oscillation contrast and the population overlap of the 4-qubit GHZ state, and compare with our error model predictions, assuming perfect CZ gates. (Ext. Data Fig.~\ref{SI_ZNE_data}).

The experimental pulse sequence is shown in Ext. Data Fig.~\ref{SI_ZNE_data}a. We increase the total idle time from 280 $\mathrm{\mu s}$ to 840 $\mathrm{\mu s}$ per arm and observe a decrease in both GHZ state population overlap and parity oscillation contrast (Ext. Data Fig.~\ref{SI_ZNE_data}c,e). We also observe similar trends in numerical simulation with our error model, assuming perfect CZ gates, finite temperature of $\Bar{n}=0.24$, and the calibrated clock laser frequency PSD. 

With the actual reconfiguration time ($280~\mathrm{\mu s}$), this error model predicts the parity oscillation contrast to be 0.66 and the state fidelity to be 0.75, consistent with our experimental realization (contrast being 0.68(3) and fidelity being 0.71(2)).

\subsubsection*{Error model with a 26-mHz clock laser system}
The experimental results from variable idle time have shown that the 4-qubit GHZ state generation fidelity is limited by the clock laser frequency noise. This motivates us to simulate this state generation circuit with our clock error model and the frequency PSD of a 26-mHz laser\cite{Bishop2013}. Keeping a finite temperature of $\Bar{n}=0.24$ and assuming perfect CZ gates, we find a contrast of $0.79$ and a fidelity of $0.84$ (Ext. Data Fig.~\ref{SI_sim_better_clock}b). With this reduced frequency noise, we find the 4-qubit GHZ generation fidelity less sensitive to the idle time. For the simultaneous generation of a cascade of GHZ states, we find consistent 4-qubit GHZ fidelity with the shorter idle time sequence.

Furthermore, with zero temperature ($\Bar{n}=0$), the clock error model predicts near-unity state fidelity ($>0.999$). In this low-temperature and 26-mHz clock laser scenario, the state fidelity is limited by the entangling gate fidelity. With the high-fidelity entangling gate demonstrated in this work, we estimate the 4-qubit GHZ state generation fidelity to be $\approx0.97$.

This improvement of atomic temperature could readily be achieved by erasure cooling~\cite{Scholl2023B} or other methods~\cite{Norcia2018}. We note that erasure-cooling was not applied during this particular experiment for faster data-taking on a 4-atom register.

\subsection*{Projected metrological gain}
We analyze the experimental fidelities required to obtain a metrological gain in phase estimation. The metrological gain $g$ is defined as the ratio of posterior variances~\cite{Pezze2018}. If we consider the gain of a protocol with $N$ entangled atoms over the interrogation of $N$ uncorrelated atoms, it can be written as $g = (\Delta\phi_{UC})^2/(\Delta\phi_{C})^2$, where $(\Delta\phi_{C})^2$, $(\Delta\phi_{UC})^2$ are the posterior variances for the entangled case and the uncorrelated case, respectively. For both cases, we assume dual-quadrature readout~\cite{Rosenband2013, Shaw2024, Zheng2024}. We first describe the expected metrological gain with perfect state preparation and then consider the case of imperfect state preparation. 

There are two distinct regimes for phase estimation. Local phase estimation corresponds to the limit of a vanishing prior phase width, or equivalently for short interrogation times in atomic clocks. This limit holds only if the prior phase width is smaller than the dynamic range of the quantum state. In this limit, the optimal probe state is an $N$-atom GHZ state, and the gain is~\cite{Pezze2018} $g \approx N$.

For the case of a large phase prior distribution width, or equivalently for long interrogation times in atomic clocks, GHZ states do not provide a metrological gain due to their limited dynamic range, which requires finding new protocols. In the main text, we consider the protocol proposed in Refs.~\cite{Giovannetti2006, Kessler2014A} and demonstrate a scheme to generate the required input state and read out the phase in both quadratures (Fig.~\ref{Fig_GHZ}). The protocol utilizes $N$ atoms divided into $M$ groups of GHZ states with $K=2^{j}$ numbers of atoms each, where $j=0,..,M-1.$ The number of atoms in the largest GHZ state is thus $K_{\text{max}}=2^{M-1}.$ The projected metrological gain with ideal state preparation was predicted to be~\cite{Kessler2014A} $g\approx \pi^2 N/(64\log(N))$. The gain can be understood by considering phase estimation at the Heisenberg limit for the largest GHZ state which contains $N/2$ atoms. To exponentially suppress rounding errors in phase estimation, one needs to use $n_0$ copies of each GHZ state with $n_0=(16/\pi^2)\mathrm{log(N)}$. We note that these expressions only hold in the limit of large $N$.

We now consider a limited number of atoms, $N$, and analyze the effect of finite state preparation fidelity on the projected metrological gain. We assume that the interrogation time is long, and perform numerical Bayesian phase estimation. We use a Gaussian prior distribution to model the prior knowledge of the laser phase. For Bayesian phase estimation, the posterior variance of a given protocol depends on the variance of the prior distribution. The figure of merit of phase estimation is typically given by~\cite{Macieszczak2014,Kaubruegger2021}: $R=\Delta\phi_{C}/\delta\phi$, where $\delta \phi$ is the prior phase distribution width. $R$ quantifies how much information was obtained in the measurement compared to our initial knowledge of the parameter. It was shown numerically that for relevant values of $N$ ($N \approx 100$), the optimal performance, quantified by $R,$ is obtained for a phase prior width around $\delta\phi = 0.7$ rad for a large class of states~\cite{Macieszczak2014,Kaubruegger2021}. We therefore choose to work with this prior width. Experimentally, the prior width is set by the Ramsey interrogation time and can be tuned to this optimal value. 

\begin{figure*}[]
	\centering
	\includegraphics[width=\textwidth]{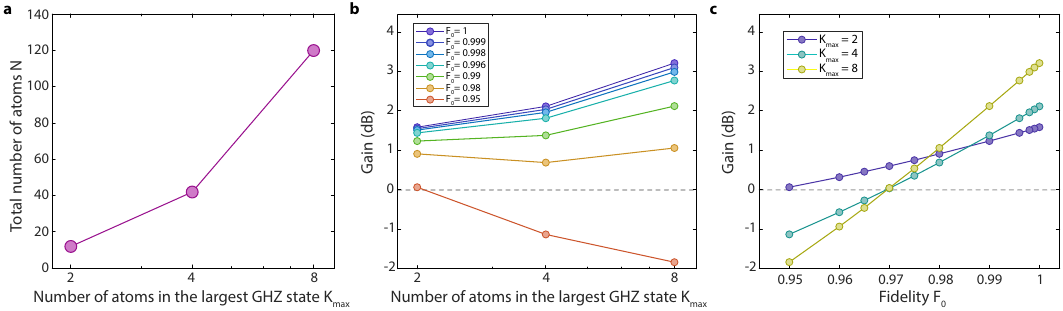}
	\caption{\textbf{Projected metrological gain.} \textbf{a}, Total number of atoms $N$ required as a function of the number of atoms in the largest GHZ state used in the protocol proposed in Ref. \cite{Kessler2014A}. Efficient implementation requires $n_0$ copies of each GHZ state with $2^j$ atoms, $j = 0, 1, \dots M-1$, where the largest GHZ state contains $K_{max} = 2^{M-1}$ atoms, and $N = n_0(2^{M}-1)$. The number of copies for $K_{max} = 2, 4, 8$ are $n_0 = 6, 8, 9$ respectively.  \textbf{b}, Projected metrological gain for various effective fidelities per qubit $F_0$, as a function of the number of atoms in the largest GHZ state, $K_{max}$. We assume the contrast $C$ of the parity oscillations for a $K$-atom GHZ state to scale as $C(K) = F_0^K$. \textbf{c}, Projected metrological gain as a function of effective fidelity per qubit $F_0$, for maximum GHZ size $K_{max} = 2, 4, 8$. We see that the effect of finite fidelity is more prominent for states with larger maximum GHZ sizes, causing the gain to decrease steeply with decreasing fidelity.
	} 

	\label{SI_metrological_gain}
\end{figure*}

\begin{figure*}[htb!]
	\centering
	\includegraphics[width=\columnwidth]{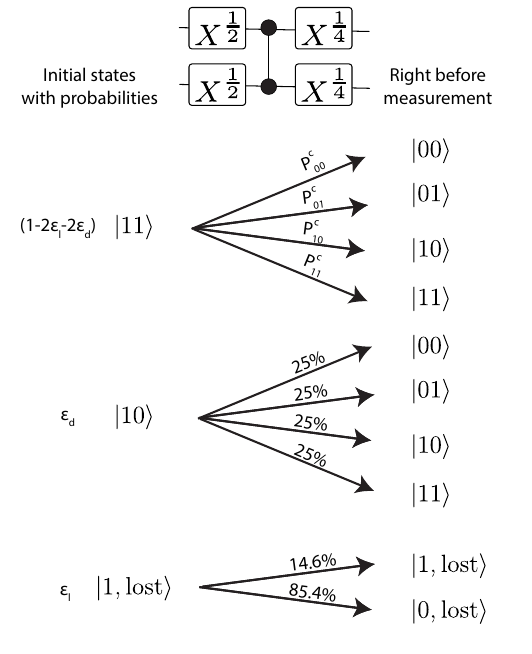}
	\caption{\textbf{State-preparation correction for Bell state population.} We draw the probability tree, starting from circuit initialization (state preparation), and track the evolution up to the measurement stage. From this, we can correct for state preparation errors. Following erasure-cooling as part of the state preparation, we have the ideal initial state and erroneous initial states with respective probabilities. We track their evolution under the Bell state generation circuit and see how each initial state contributes to the population distribution before the measurement. The SPAM corrected values, labeled by $P_b^c$, where $b$ labels the different two-qubit bitstring, are the quantities that we want to evaluate. The pathway for $\ket{\text{lost}, 1}$ and $\ket{01}$ can be inferred easily from the clear symmetry with $\ket{1, \text{lost}}$ and $\ket{10}$. For simplicity, $\ket{00}, \ket{\text{lost}, \text{lost}}, \ket{0, \text{lost}}, \ket{\text{lost}, 0}$ are not shown due to their small probability on the order of $\varepsilon_d^2$, $\varepsilon_l^2$ $\approx 10^{-5}$.
    }
	\label{SI_Bell_SPAM}
\end{figure*}

\begin{figure*}[]
	\centering
	\includegraphics[width=\textwidth]{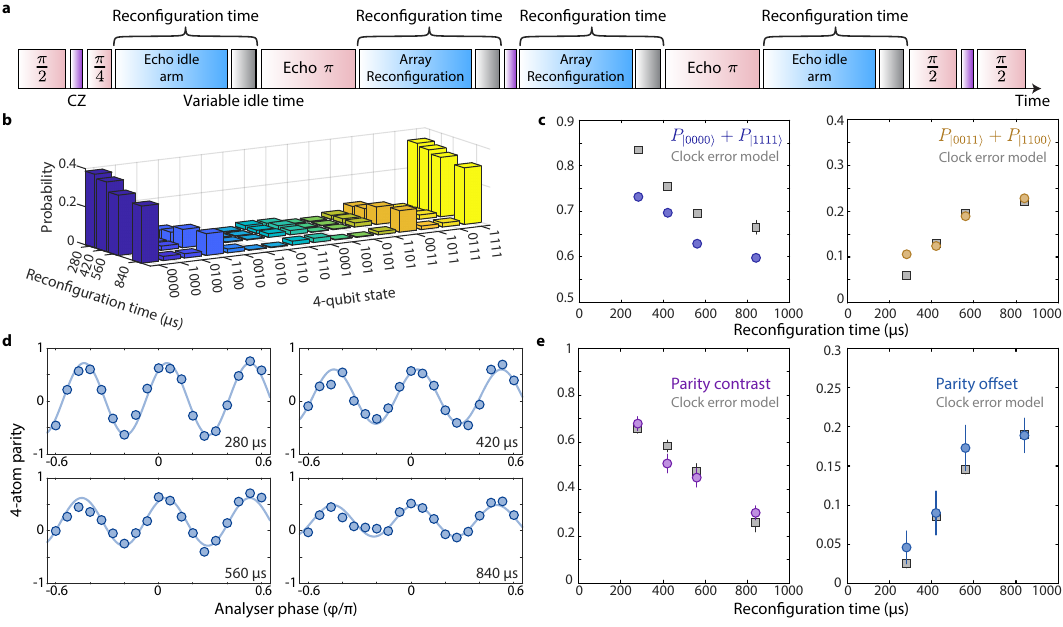}
    \caption{\textbf{Atom-laser dephasing effect on 4-qubit GHZ state preparation.} \textbf{a}, Experimental pulse sequence for the circuit described in Fig.~\ref{Fig_GHZ}a. In the actual sequence, echo pulses are added during the array reconfiguration operation, which are not shown in the main text for simplicity. The relative sizes of the blocks represent the relative duration of each operation, except for the Rydberg CZ gate (which takes a much shorter time) and the variable idle time (grey blocks) which is adjusted to vary the array reconfiguration time (total idle time) to study the effect due to clock laser frequency noise, or equivalently atom-laser dephasing. In this sequence, $\pi$ pulse takes $\approx$ 240 $\mu$s, and the CZ gates take $\approx$ 500 ns. The actual array reconfiguration (blue blocks) takes 280 $\mu$s. \textbf{b}, Experimental raw data of the population distribution as a function of total reconfiguration time. \textbf{c}, We plot the target state population overlap and the growth of dominant non-target states ($\ket{0011}$ and $\ket{1100}$) as a function of total reconfiguration time. Our error model predictions (grey squares) successfully verify the decreasing trend for target state population overlap and the non-target state population growth. \textbf{d}, Experimental raw data of the parity oscillation contrast for different total reconfiguration times. \textbf{e}, Decay of parity contrast and growth of parity offset. Our error model predictions (grey squares) are in good agreement with experimental results (see Methods for error model details).
    }  
	\label{SI_ZNE_data}
 \end{figure*}

Using the protocol in Ref.~\cite{Kessler2014A} for GHZ states with 1, 2, and 4 atoms, the minimal number of copies per GHZ size is $n_0 = 6$, resulting in a total number of $N = 42$ atoms (Ext. Data Fig.~\ref{SI_metrological_gain}). Using optimal Bayesian estimators~\cite{Macieszczak2014}, we numerically find a metrological gain of 1.627 (2.114 dB) with perfect state preparation. Imperfect preparation fidelity results in a parity signal with limited contrast $C(K)$ for a $K$-atom GHZ state. The probabilities for the outcome of a parity measurement are then modified to \mbox{$P(\pm) = (1 \pm C(K)\cos{(K\phi)})/2$}, where $\{ +, -\}$ denote even, odd parity.

To estimate the effect of such limited contrast on the metrological gain we consider two characteristic scenarios, motivated by our experimental results. First, we look at a case with perfect state preparation for the 1-atom and 2-atom GHZ states ($C(1)=C(2)=1$), while the 4-atom GHZ state has a finite parity signal contrast $C(4)$.In this case, we numerically find the minimal contrast to obtain a gain $g \geq 1$ is $C(4) = 0.656$. This is higher than the threshold contrast for a narrow prior width given the same state, i.e. $C \geq 1/\sqrt{K_{\text{max}}}=0.5$~\cite{Pezze2018, Colombo2022}.

Second, we look at a case where the contrast of each GHZ state scales as $C(K)=F_0^K$, where $F_0$ is an effective fidelity per qubit. We repeat the calculation and find the threshold fidelity to obtain a gain to be $F_0 = 0.969$, or equivalently a contrast of $C(4) = 0.969^4 = 0.883$ for the 4-atom GHZ state. Numerically, this threshold seems to be robust to the introduction of additional copies of the 1-atom and 2-atom GHZ states: if $n_0 = 12$ only for these states (keeping $n_0 = 6$ for the 4-atom GHZ states), the threshold is only slightly reduced to $F_0 \gtrsim 0.965$. Finally, we show the projected metrological gain for various $F_0$, with respect to the number of atoms in the largest GHZ state used in the protocol of Ref.~\cite{Kessler2014A} in Ext. Data Fig.~\ref{SI_metrological_gain}.

Our experimental values for $C(K)$ from the simultaneous GHZ state generation scheme ($C(1) = 0.82, C(2) = 0.68, C(4) = 0.52$) fall between the two cases considered here: the best case scenario of $C(K)=1$ for $K<K_{\text{max}}$, and the worst case scenario of $C(K)=F_0^K$. Hence we expect the threshold for metrological gain to lie between the values obtained from these two cases.

We note that the observed parity oscillation contrasts in our current experimental demonstration are below these thresholds. However, the contrast reduction is entirely dominated by clock laser noise (Ext. Data Fig. ~\ref{SI_sim_better_clock}). With the high-fidelity CZ gates obtained in this work, where $F_{CZ} \approx 0.996$, combined with a reduced clock laser noise (achieved for example by a laser with frequency PSD similar to Ref.~\cite{Bishop2013} as discussed above), we numerically project a performance superior to the same number of uncorrelated atoms. Specifically, if we assume $F_0\approx 0.996$, the predicted metrological gain with 6 copies of GHZ states of 1, 2, and 4 atoms each is calculated to be 1.519 (1.815 dB). If 8-atom GHZ states are included, the predicted metrological gain with 8 copies of GHZ states with 1, 2, 4, and 8 atoms each is calculated to be 1.893 (2.772 dB).
We remark that the above gain analysis assumes zero dead time. Introducing dead time would degrade the gain and we defer analysis of this effect to a future work.

\subsection*{Repeated ancilla detection with ancilla reuse}
\begin{figure*}[]
	\centering
	\includegraphics[width=\columnwidth]{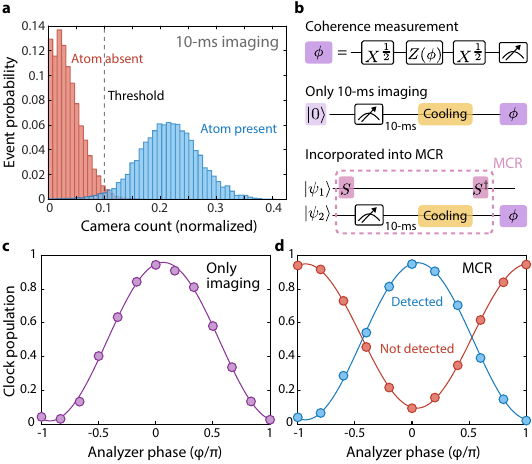}
    \caption{\textbf{High-survival mid-circuit imaging with ancilla recool and reuse.} \textbf{a}, Histograms of the normalized camera count for 10-ms imaging. Using 120-ms standard imaging, we can detect if a tweezer is empty (red) or filled (blue) with $\approx 0.9999$ fidelity prior to the 10-ms imaging. In this scheme, the typical detection fidelity which corresponds to equal error probability in detecting absence or presence of an atom is 0.98. Survival after imaging is 0.965(2), as opposed to the lossy 18 $\mu$s fast imaging employed in the main text (Fig.~\ref{Fig_ancilla_parity}). \textbf{b}, We check if we can coherently apply single-qubit rotations by applying a $\pi/2$ pulse and a second $\pi/2$ pulse with a variable phase. We then extract the parity contrast as a way of coherence measurement. We first measure the coherence after this 10-ms imaging. Then we incorporate this into MCR and measure the coherence. \textbf{c}, The parity contrast of atoms after 10-ms imaging and cooling is measured to be 0.94(1). \textbf{d}, The parity contrast of atoms after being detected during the MCR reads 0.93(1) (blue) while that of the atoms not detected during the MCR reads 0.86(1) (red). The difference can be explained by the finite decay from $^3\text{P}_0 (\ket{1})$ to $^1\text{S}_0 (\ket{0})$ (see Methods).
    }  
\label{SI_ancilla_reset}
\end{figure*}

\begin{figure*}[]
	\centering
	\includegraphics[]{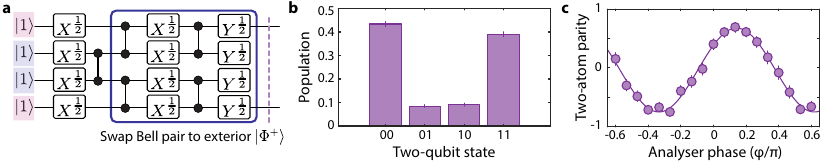}
	\caption{\textbf{State exchange operations of a Bell pair between two quantum registers.} \textbf{a}, As part of the ancilla-based parity readout (Fig.~\ref{Fig_ancilla_parity}a,b), we employ a state exchange between two quantum registers. Shown is the generation and swap of a Bell pair from two inner atoms to two outer atoms which did not interact directly. \textbf{b, c}, Characterization of the resulting long-range Bell pair. We obtain a population overlap of 0.842(8) and a parity contrast of 0.71(2), yielding a raw fidelity of 0.78(1).
	} 
	\label{SISwap_Bell}
\end{figure*}

For mid-circuit readout performed in Fig.~\ref{Fig_ancilla_single}, fast 18 $\mu$s imaging~\cite{Scholl2023A} is applied with a fidelity of $\approx$ 0.96 at tweezer spacing of 3.3 $\mu$m. The strong driving on the $^1\text{S}_0$ $\leftrightarrow$ $^1\text{P}_1$ transition without cooling results in low survival of detected atoms. Therefore, for experiments where repeated ancilla measurement is needed, we refill the original ancilla position with another atom via array reconfiguration. Alternatively, we can recool and reuse the ancilla atoms instead of refilling them. In this section, we show a proof-of-concept experiment with different imaging parameters on the ancilla atoms (Ext. Data Fig.~\ref{SI_ancilla_reset}).

The new imaging scheme is based on the standard high-fidelity, high-survival imaging with cooling light ( $^1\text{S}_0$ $\leftrightarrow$ $^3\text{P}_1$ intercombination line)~\cite{Covey2019A}. We increase the imaging power to collect more photons for 10 ms and apply the cooling light on one of the axes for another 10 ms, shorter than the motional shelving coherence time $\approx$ 100 ms~\cite{Scholl2023B}. This imaging scheme allows us to obtain an imaging fidelity of 0.98 with 0.965(2) survival (Ext. Data Fig.~\ref{SI_ancilla_reset}a).

We then check if we can coherently apply single-qubit rotations after this 10-ms imaging by applying a $\pi/2$ pulse and a second $\pi/2$ pulse with a variable phase (Fig.~\ref{SI_ancilla_reset}b). The measured coherence after imaging 0.94(1) (Fig.~\ref{SI_ancilla_reset}c) is mainly limited by survival which can be readily improved with further optimizations on cooling during the imaging. Once added to the complete MCR (Fig.~\ref{SI_ancilla_reset}d), we see similar coherence for the detected atoms in the ground state (blue) and a slightly lower coherence for the non-detected atoms in the clock state (red), due to the decay (time constant $\approx$ 300~ms) during the 10-ms cooling. These decayed atoms contribute doubly to the loss of coherence $\approx0.07$. With these coherent driving, we see that the ancilla atoms are ready to be reused. In the same experiment, we also measured a coherence of $\approx$ 0.73 on the shelved atoms after unshelving them (not shown), matching the numbers of motional coherence in our previous work~\cite{Scholl2023B}.

\subsection*{Weight-2 ancilla-based parity readout}
We give the fitted parameters of the plot presented in Fig.~\ref{Fig_ancilla_parity}b. For the direct readout on the Bell pair, we fit an oscillation of 16.5(1) kHz with a phase 3.16(7) rad and a contrast of 0.77(3). Ancilla readout gives a Ramsey oscillation of 16.5(2) kHz with a phase of 3.19(9) rad and a contrast of 0.59(4). In a separate experiment interleaved with this experiment, we measured single-atom detuning to be 8.26(5) kHz.

With perfect gates, the quantum circuit in Fig.~\ref{Fig_ancilla_parity}a yields an oscillating state between $\ket{\Phi^+}\otimes\ket{0}_{\text{ancilla}}$ and $\ket{\Phi^-}\otimes\ket{1}_{\text{ancilla}}$. For either state, the pair would be measured in $\ket{00}$ or $\ket{11}$ in this ideal case. Given this information, one can post-select on the experimental repetitions where the pair is measured in the expected state, either $\ket{00}$ or $\ket{11}$. This post-selection could identify errors in the execution of the circuit. Looking at the oscillating state, this post-selection should not bias either of the ancilla measurement outcomes. Performing this post-selection analysis (not shown), we see a Ramsey oscillation of 16.5(3) kHz and a contrast of 0.71(8).


\end{document}